%% file: main_17.4.tex
\documentclass[aps,prb,reprint,superscriptaddress,floatfix,footinbib,nobibnotes]{revtex4-1}

\usepackage{amsmath}
\usepackage{amssymb}
\usepackage{graphicx}
\usepackage[separate-uncertainty = true,multi-part-units=single]{siunitx}
\usepackage{natbib}
\usepackage{chapterbib}
\usepackage{hyperref}
\usepackage{array}
\usepackage[utf8]{inputenc}
\usepackage[english]{babel}

\usepackage{comment}

\newcommand{\sgn}{\operatorname{sgn}}
\newcommand{\lin}[1]{\mathrm{#1}}
\newcommand{\sigmaO}{\sigma_\lin{o}^\lin{SOT}}
\newcommand{\sigmaE}{\sigma_\lin{e}^\lin{SOT}}
\newcommand{\sigmaF}{\sigma_\lin{e}^\lin{F}}

\newcommand{\Ms}{M_\lin{s}}
\newcommand{\Msd}{M_\lin{s} d_\lin{FM}}
\newcommand{\Py}{Ni$_{80}$Fe$_{20}$}

\DeclareSIUnit\torr{Torr}
\DeclareSIUnit\sq{\ensuremath{\Box}}

\begin{document}


\title{Inductive detection of field-like and damping-like AC inverse spin-orbit torques in ferromagnet/normal metal bilayers}

\author{Andrew J. Berger}
\affiliation{Quantum Electromagnetics Division, National Institute of Standards and Technology, Boulder, CO 80305, U.S.A.}
\thanks{Contribution of the National Institute of Standards and Technology; not subject to copyright.}

\author{Eric R. J. Edwards}
\affiliation{Quantum Electromagnetics Division, National Institute of Standards and Technology, Boulder, CO 80305, U.S.A.}

\author{Hans T. Nembach}
\affiliation{Quantum Electromagnetics Division, National Institute of Standards and Technology, Boulder, CO 80305, U.S.A.}


\author{Alexy D. Karenowska}
\affiliation{Department of Physics, University of Oxford, Oxford, U.K.}

\author{Mathias Weiler}
\affiliation{Walther-Mei{\ss}ner-Institut, Bayerische Akademie der Wissenschaften, Garching, Germany}
\affiliation{Physik-Department, Technische Universität München, Garching, Germany}

\author{Thomas J. Silva}
\email[]{thomas.silva@nist.gov}
\affiliation{Quantum Electromagnetics Division, National Institute of Standards and Technology, Boulder, CO 80305, U.S.A.}

\date{\today}

\begin{abstract}
Functional spintronic devices rely on spin-charge interconversion effects, such as the reciprocal processes of electric field-driven spin torque and magnetization dynamics-driven spin and charge flow. Both damping-like and field-like spin-orbit torques have been observed in the forward process of current-driven spin torque and damping-like inverse spin-orbit torque has been well-studied via spin pumping into heavy metal layers. Here we demonstrate that established microwave transmission spectroscopy of ferromagnet/normal metal bilayers under ferromagnetic resonance can be used to inductively detect the AC charge currents driven by the inverse spin-charge conversion processes. This technique relies on vector network analyzer ferromagnetic resonance (VNA-FMR) measurements. We show that in addition to the commonly-extracted spectroscopic information, VNA-FMR measurements can be used to quantify the magnitude and phase of all AC charge currents in the sample, including those due to spin pumping and spin-charge conversion. Our findings reveal that \Py/Pt bilayers exhibit both damping-like and field-like inverse spin-orbit torques. While the magnitudes of both the damping-like and field-like inverse spin-orbit torque are of comparable scale to prior reported values for similar material systems, we observed a significant dependence of the damping-like magnitude on the order of deposition. This suggests interface quality plays an important role in the overall strength of the damping-like spin-to-charge conversion.
\end{abstract}

\maketitle

\include{Inductive_detection_of_AC_charge_currents_v17.4_MT}
\include{Inductive_detection_of_AC_charge_currents_v17.4_SI}


\end{document}

%% file: Inductive_detection_of_AC_charge_currents_v17.4_MT.tex
\section{Introduction}

Spin-charge transduction effects for ferromagnet/nonmagnet (FM/NM) multilayers couple electric fields to magnetic torques in the forward process (so-called spin-orbit torque (SOT)), and they couple magnetization dynamics to currents in the inverse process (iSOT). In general, these torques can be phenomenologically separated into two components: damping-like and field-like. Both are perpendicular to the FM magnetization, but the damping-like torque is odd under time-reversal and dissipative, whereas the field-like torque is even under time-reversal and conservative \cite{freimuth_direct_2015}. A classic example of a field-like torque is the action of an Oersted field on a FM magnetization due to a charge current in an adjacent conducting layer. By Onsager reciprocity, the inverse process is captured by Faraday's law: magnetization dynamics in the FM generate charge currents in the NM. Recently, it has been found that spin-orbit coupling (SOC) in multilayers can give rise to both field- and damping-like SOTs \cite{mihai_miron_current-driven_2010,duan_nanowire_2014}, but with substantially different scaling than that achieved with Oersted fields. Unlike the Oersted effect, these spin-orbitronic effects are short-range, making them highly advantageous for microelectronic applications that require device scaling to high densities such as nonvolatile memory and alternative state-variable logic \cite{liu_spin-torque_2012, manchon_new_2015}.

Damping-like torques due to the spin Hall effect (SHE) in heavy NM layers such as Pt and $\beta$-Ta are well-studied and understood, and have been investigated in both forward \cite{liu_spin-torque_2012} and inverse configurations \cite{czeschka_scaling_2011, weiler_detection_2014, wang_scaling_2014}. Substantial field-like torques have also been measured for FM/NM interfaces in the forward configuration \cite{mihai_miron_current-driven_2010, garello_symmetry_2013, avci_fieldlike_2014, fan_quantifying_2014}. However, an inverse measurement of the field-like torque in \Py/Pt has not yet been unambiguosly reported \cite{weiler_phase-sensitive_2014}. Here, we present simultaneous measurements of inverse field-like and damping-like torques in \Py/Pt bilayers via well-established coplanar waveguide (CPW) ferromagnetic resonance (FMR). Time-varying magnetic fields produced by a FM/NM sample under FMR excitation will inductively couple into the CPW, altering the total inductance of the microwave circuit. Such fields are produced by: (1) the Py precessing magnetization, (2) Faraday effect induced AC currents in the Pt layer, and (3) spin-orbit AC currents due to damping-like and (4) field-like processes. We show that through proper background normalization, combined with Onsager reciprocity for the specific phenomenology of these measurements, commonly-used vector network analyzer (VNA) FMR spectroscopy allows accurate identification of the processes that contribute to spin-charge conversion. 

The paper is organized as follows. In Sec. \ref{sec:Theory}, by appealing to Onsager reciprocity we provide the phenomenological background relating the forward and inverse processes that produce magnetic torques and charge flow in a ferromagnet/normal metal system under electrical bias or with excited magnetization dynamics. Sec. \ref{sec:ExpTech} describes the quantitative VNA-FMR technique, and derives the expressions we use to calculate the sample's complex inductance. This section also introduces the effective conductivity $\tilde{\sigma}_\lin{NM}$ that quantifies the magnitude and symmetry of magnetic torques due to applied charge currents, and reciprocally, of the AC charge currents flowing in a sample in response to the driven magnetization dynamics. In Sec. \ref{sec:Data}, we present data acquired from \Py/Pt bilayers and \Py/Cu control samples. The magnitude of the phenomenological parameter $\tilde{\sigma}_\lin{NM}$ extracted from these data is well within the range of reported values, and it obeys the usual symmetry properties associated with the stacking order of the \Py and Pt layers. Finally, we discuss the results in Sec. \ref{sec:Discuss} by comparing our extracted iSOT parameters to the microscopic spin-charge conversion parameters of spin Hall angle and Rashba parameter. In all cases, the magnitudes of the extracted spin Hall angle and Rashba parameter are in rough agreement with what has been reported in the literature, though this agreement is contingent on the assumption of typical values for the interfacial and bulk spin transport parameters. However, we find that the extracted spin Hall angle changes by a factor of almost 4 depending on the growth order of the multilayer stacks, with a larger spin Hall angle when the Pt is grown on top of the \Py. This suggests that the spin transport parameters are in actuality highly dependent on the stack growth order.


\section{Onsager relations for spin-orbit torque}
\label{sec:Theory}
Onsager reciprocity relations \cite{onsager_reciprocal_1931} are well known for certain pairs of forces and flows. For example, for thermoelectric effects, applied electric fields or thermal gradients can drive both charge and heat flow. In this section, we establish Onsager relations for charge current and magnetic torque as the flows that are driven by applied electric fields and magnetization dynamics in a FM/NM multilayer \cite{freimuth_direct_2015}. 

By analogy to Ohm's Law, $\mathbf{J} = \sigma \mathbf{E}$, we can write a general matrix equation relating driving forces (magnetization dynamics $\partial\hat{m}/\partial t$ and electric field $\mathbf{E}$) to flows (magnetic torque density $\mathbf{T}$ and charge current density $\mathbf{J}$) \cite{freimuth_direct_2015}:

\begin{widetext}
\begin{equation}
\begin{split}
&\begin{bmatrix}
\left(\dfrac{2e}{\hbar}\right) \left[ \displaystyle\int\limits_0^{+d_\lin{FM}} \mathbf{T}(z) dz \right] \\
\left[\displaystyle\int\limits_{-d_\lin{NM}}^{+d_\lin{FM}} \mathbf{J}(z) dz\right]
\end{bmatrix} = \\
\mathcal{G}
&\begin{bmatrix}
G_\lin{mag} && \sgn(\hat{z}\cdot\hat{n})\left( -\sigmaF + \sigmaE - \sigmaO [\hat{m}\times]\right) \\
\sgn(\hat{z}\cdot\hat{n})\left( -\sigmaF + \sigmaE - \sigmaO [\hat{m}\times]\right) && -\dfrac{1}{Z_\lin{eff}}
\end{bmatrix} \\
&* \begin{bmatrix}
\left(\dfrac{\hbar}{2e}\right)\dfrac{\partial  \hat{m}}{\partial t} \\
\hat{z} \times \mathbf{E}
\end{bmatrix}
\label{eq:OnsagerMatrix}
\end{split}
\end{equation}
\end{widetext}

\noindent where $\hat{m}$ is the magnetization unit vector, $\hbar$ is Planck's constant divided by $2 \pi$, $e$ is the electron charge, $d_\lin{FM}$ and $d_\lin{NM}$ are the FM and NM thicknesses. The terms in the $2 \times 2$ conductivity matrix are described below. The sign of the off-diagonal terms are determined by $\sgn(\hat{z}\cdot\hat{n})$, where $\hat{n}$ is an interface normal pointing into the FM. The coordinate unit vector $\hat{z}$ is defined by the sample placement on the CPW, as shown in Fig. \ref{fig:CPWandCoords}(a), and $z=0$ is defined by the FM/NM interface. $\mathcal{G}$ is a $2 \times 2$ matrix imposing geometrical constraints: (1) magnetic torques are orthogonal to $\hat{m}$ and (2) charge currents can flow only in the $x,y$ plane:

\begin{equation}
\mathcal{G} = 
\begin{bmatrix}
[\hat{m}\times] && 0 \\
0 && [\hat{z}\times] 
\end{bmatrix}
\end{equation}

The diagonal elements of the effective conductivity matrix describe the Gilbert damping of the FM and charge flow in the metallic bilayer in response to an applied electric field. That is,

\begin{eqnarray}
\left(\dfrac{2e}{\hbar}\right) \left[ \displaystyle\int\limits_0^{+d_\lin{FM}} \mathbf{T}(z) dz \right] &=& \left(\dfrac{\hbar}{2e}\right)  G_\lin{mag} \left( \hat{m} \times \dfrac{\partial  \hat{m}}{\partial t}\right) \label{eq:GilbertTorque} \\
\left[\displaystyle\int\limits_{-d_\lin{NM}}^{+d_\lin{FM}} \mathbf{J}(z) dz\right] &=& -\dfrac{1}{Z_\lin{eff}} \hat{z} \times ( \hat{z} \times \mathbf{E} ) \label{eq:OhmsLaw}
\end{eqnarray}

\noindent where $G_\lin{mag} \equiv -d_\lin{FM}(2e/\hbar)^2 (\alpha M_\lin{s}/\gamma)$, $\alpha$ is the Gilbert damping parameter, and $\gamma$ is the gyromagnetic ratio, such that Eq. \ref{eq:GilbertTorque} is the usual Gilbert damping term from the Landau-Lifshitz-Gilbert equation: 

\begin{equation}
\frac{\partial \hat{m}}{\partial t} = -\gamma \mu_0 \hat{m} \times \mathbf{H} - \left(\frac{\gamma}{M_\lin{s}d_\lin{FM}}\right) \int\limits_0^{+d_\lin{FM}} \mathbf{T}(z) dz
\label{eq:LLG}
\end{equation}

In Eq. \ref{eq:OhmsLaw}, $Z_\lin{eff}$ is the effective frequency-dependent impedance of the bilayer. Eq. \ref{eq:OhmsLaw} reduces to Ohm's Law in the DC limit ($Z_\lin{eff} \rightarrow R_{\Box}$ as $f \rightarrow 0$).

The off-diagonal terms describe the electromagnetic reciprocity between Faraday's and Ampere's Law \cite{ramo_fields_2008, white_introduction_1985}, as well as spin-orbit torques (SOT) and their inverse, using the effective conductivities $\sigmaF$, $\sigmaE$, and $\sigmaO$. 

\begin{widetext}
\begin{eqnarray}
\left(\dfrac{2e}{\hbar}\right) \left[ \displaystyle\int\limits_0^{+d_\lin{FM}} \mathbf{T}(z) dz \right] &=& \sgn(\hat{z}\cdot\hat{n}) \hat{m} \times \left( -\sigmaF + \sigmaE - \sigmaO [\hat{m}\times]\right) (\hat{z}\times\mathbf{E}) \label{eq:forwardSOT} \\
\left[\displaystyle\int\limits_{-d_\lin{NM}}^{+d_\lin{FM}} \mathbf{J}(z) dz\right] &=& \sgn(\hat{z}\cdot\hat{n}) \left(\frac{\hbar}{2e}\right)\hat{z} \times \left( -\sigmaF + \sigmaE - \sigmaO [\hat{m}\times]\right) \dfrac{\partial  \hat{m}}{\partial t} \label{eq:inverseSOT} 
\end{eqnarray}
\end{widetext}

\noindent Here, the superscripts indicate the source of the torque or current as due to the Faraday effect or SOT. The subscripts indicate ``even'' or ``odd'' with respect to time-reversal, which determines the torque direction or phase of the corresponding current with respect to the driving electric field or magnetization dynamics. 

First consider the Faraday conductivity, $\sigmaF$. In the forward process an electric field $\mathbf{E}$ produces a charge current, which by Ampere's Law produces a magnetic field. This field exerts a torque $\mathbf{T}$ on the magnetization of the FM layer. In the reverse process, magnetization dynamics $\partial_t \hat{m}$ produce an AC magnetic field, which by Faraday's Law induces a charge current $\mathbf{J}$ in the NM layer. In this way, $\sigmaF$ quantifies the reciprocity between Ampere's and Faraday's Law (see Eq. \ref{eq:sigmaF} for an estimate of the $\sigmaF$ magnitude based on material properties). Inclusion of the terms in Eq. \ref{eq:OnsagerMatrix} due to electrodynamic reciprocity is critical for the proper interpretation of inverse spin orbit torque experiments \cite{weiler_phase-sensitive_2014}. 

Also present in the off-diagonal terms are SOT conductivities due to spin-charge conversion. In Eq. \ref{eq:forwardSOT}, these manifest as electric-field driven damping-like torques, which are proportional to $\hat{m} \times(\hat{m} \times (\hat{z} \times \mathbf{E}))$, and field-like torques, which are proportional to $\hat{m} \times(\hat{z} \times \mathbf{E})$. The constants of proportionality between applied electric field and SOTs are $\sigmaO$ and $\sigmaE$. In the reverse direction (Eq. \ref{eq:inverseSOT}), these effects are responsible for spin-to-charge conversion (e.g., inverse spin Hall effect (iSHE) \cite{saitoh_conversion_2006} or inverse Rashba-Edelstein effect (iREE) \cite{sanchez_spin--charge_2013}). 

Reporting effective conductivities, as opposed to spin-charge conversion parameters like the spin Hall angle, directly relates the microwave inputs and charge current outputs of an iSOT device without the need for separate characterization of spin-mixing conductance or spin diffusion length. Reciprocally, in a spin torque experiment with charge current inputs and magnetization dynamics (or switching) as output, the effective conductivities provide the ideal figure of merit for determining magnetization oscillation and switching thresholds of the applied current. To estimate the critical current density $J_\lin{c}$ needed to switch the magnetization of a ferromagnetic layer \cite{katine_current-driven_2000, mangin_current-induced_2006}, one simply needs to equate the Gilbert damping torque (Eq. \ref{eq:GilbertTorque}) and odd (anti-damping-like) SOT (Eq. \ref{eq:forwardSOT}): 

\begin{equation}
J_\lin{c} = \alpha \Msd \frac{\omega}{\gamma} \frac{2e}{\hbar} \left(\frac{\sigma}{\sigma_\lin{o}^\lin{SOT}}\right)
\end{equation}

\noindent where $\omega$ is the FMR frequency with no applied fields (e.g. for in-plane magnetization, $\omega = \mu_0 \gamma \sqrt{H_\lin{k}(M_\lin{s} + H_\lin{k})}$, with anisotropy field $H_\lin{k}$). Using $\alpha$ as determined for these \Py/Pt films (see SI), $M_\lin{s} = $ \SI{700}{\kilo\ampere/\meter}, $H_\lin{k} = $ \SI{160}{\kilo\ampere/\meter} (for thermal stability considerations), bulk Pt resistivity \cite{lide_CRC_2003}, and the measured $\sigma_\lin{o}^\lin{SOT}$ (see Table \ref{tab:SpinChargeParams}), we estimate a critical current density of \SI{2e12}{\ampere/\m^2} for a \SI{2}{\nm} \Py film.

While the effective conductivity is the directly measured quantity, in Sec. \ref{sec:EffConductivity} we nevertheless derive expressions relating the effective conductivities to microscopic spin-charge conversion parameters. Extraction of the microscopic parameters is necessarily contingent on the details of the model employed and parameters assumed.

The effective conductivities can also be related to the often-used quantity of effective flux density per unit current density \cite{nguyen_spin_2016} $B_\lin{eff}/J$, with units of \SI{}{\tesla~ \meter^2~ \ampere^{-1}} via the equation $B_\lin{eff}/J = {\sigma}_\lin{e,o}^\lin{SOT}\hbar/(2 M_\lin{s} \sigma d_\lin{FM} e)$ (where $\sigma$ is the ordinary charge conductivity of the NM). However, our definition for the effective conductivity is more general insofar as it allows one to calculate the actual SOT without the need to independently determine the sample magnetization, conductivity, or actual thickness.

Eq. \ref{eq:OnsagerMatrix} is consistent with the phenomenological formulation presented by Freimuth, Bluegel, and Mokrousov \cite{freimuth_direct_2015}, although it has been expanded to include the purely electrodynamic contributions. Our use of the descriptors ``even'' and ``odd'' are different from that of Freimuth, et al., who use the symmetry of the spin orbit torques with respect to magnetization-reversal as the symmetry identifier. We instead use the symmetry of the torque with respect to time-reversal because this is the relevant symmetry with regard to the off-diagonal components in the phenomenological Eq. \ref{eq:OnsagerMatrix}. Any process for which the torque is odd under time-reversal qualifies as microscopically non-reversible in the sense of Onsager reciprocity, where microscopic reversibility pertains solely to forces that are even functions of velocity, as well as position \cite{onsager_reciprocal_1931}. (We also note that all axial vectors such as magnetic field are odd under time reversal.)

\section{Experimental Technique}
\label{sec:ExpTech}

\begin{figure*}[hb]
\centering
	\centering
	\includegraphics[width=\textwidth]{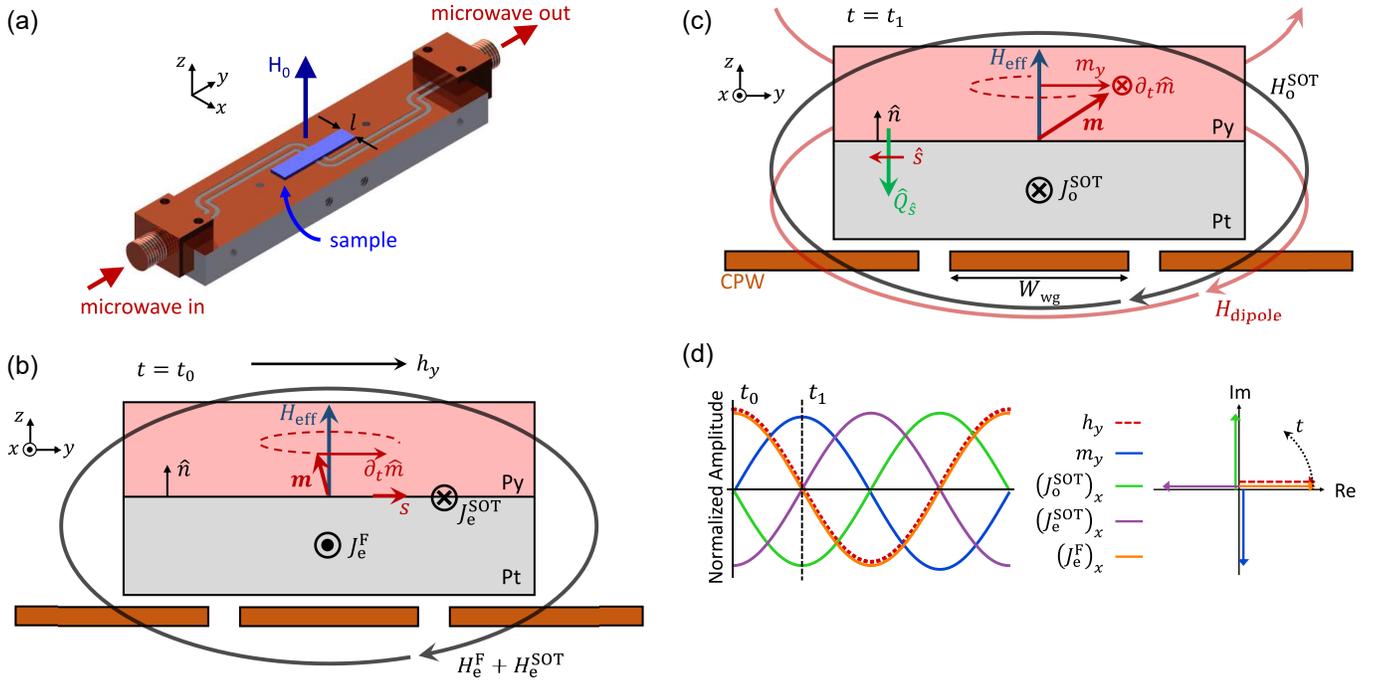}	
\caption{(a) Sample on CPW, showing out-of-plane field $H_0$ and sample length $l$. The microwave driving field points primarily along $\hat{y}$ at the sample. (b) Schematic of the bilayer, with precessing magnetization $\mathbf{m}(t)$ at time $t_0$ when $\mathbf{m} = \langle m_x,0,m_z \rangle$. Bilayer is insulated from CPW using photoresist spacer layer (not shown). At this instant in time, $J_\lin{e}^\lin{F}$ (due to the Faraday effect in the NM) and $J_\lin{e}^\lin{SOT}$ (e.g., due to inverse Rashba-Edelstein effect) are maximal along $\pm\hat{x}$, and $h_y$ is also at its maximum strength. The corresponding Oersted fields from $J_\lin{e}^\lin{F}$ and $J_\lin{e}^\lin{SOT}$ are superposed. The spin accumulation (with orientation $\hat{s}$) and $J_\lin{e}^\lin{SOT}$ are produced at the FM/NM interface. Interface normal is given by $\hat{n}$. (c) Same as (b), except at time $t_1$ when $\mathbf{m} = \langle 0,m_y,m_z \rangle$. Here, odd-symmetry SOT current $J_\lin{o}^\lin{SOT}$ (e.g., due to inverse spin Hall effect), and the dynamic fields $H_\lin{o}^\lin{SOT}$ and $H_\lin{dipole}$ are at maximum amplitude. Note that the dipolar signal is proportional to $\partial_t (H_\lin{dipole}\cdot \hat{y})$, and not simply to $H_\lin{dipole}$. Spin flow direction $\hat{Q}_{\hat{s}}$ due to spin pumping into the NM is also shown. (d) Amplitude of driving field $h_y$ and different signal sources as a function of time (left), and viewed in the complex plane at time $t_0$ (right). Relative amplitudes not indicated. For further discussion of signal phases, see SI Sec. \ref{sec:SigPhase}.}
\label{fig:CPWandCoords}
\end{figure*}

The broadband, phase-sensitive FMR measurements utilize a coplanar waveguide (CPW) as both the excitation and detection transducer (see Fig. \ref{fig:CPWandCoords}(a)). Any source of AC magnetic flux generated by the bilayer is inductively detected in the CPW. Therefore, the inductive load that the sample contributes to the CPW circuit consists of four terms: (1) The real-valued $L_0$ due to the oscillating magnetic dipolar fields produced by the resonating FM magnetization, (2) the Faraday-effect currents induced in the NM layer by the precessing FM magnetization, (3) currents produced by damping-like iSOT effects (e.g., spin pumping + iSHE), and (4) currents produced by field-like iSOT effects (e.g., iREE). The latter three inductances, which we collectively define as complex-valued $L_\lin{NM}$, are produced by currents in the NM which generate Oersted fields that inductively couple to the CPW. We quantify these currents with the effective conductivities $\sigmaF$, $\sigmaO$, and $\sigmaE$, described above. Importantly, as shown below, while $L_0$ is independent of frequency, $L_\lin{NM}$ is linear in frequency, as the currents in the NM are driven by $\partial_t \hat{m}$. Hence, frequency-dependent measurements allow us to disentangle $L_0$ and $L_\lin{NM}$. 

Figure \ref{fig:CPWandCoords}(b) and (c) show schematics of these four signal sources at two instants in time: when the dipolar and even SOT effects are maximal (Fig. \ref{fig:CPWandCoords}(b)) and when the odd SOT effect is maximal (Fig. \ref{fig:CPWandCoords}(c)). Fig. \ref{fig:CPWandCoords}(d) shows the time dependence of each of these signal sources, and their distinct phase relationships to the driving field $h_\lin{y}$, which we exploit below to determine their contributions separately.

For our measurements, we place samples onto a coplanar waveguide (CPW) with the metallic film side facing down (see Fig. \ref{fig:CPWandCoords}). This setup is positioned between the pole pieces of a room-temperature electromagnet capable of producing fields up to $\approx$\SI{2.2}{\tesla}. Using a VNA, we measure the change in microwave transmission through the CPW loaded with the bilayer sample as an out-of-plane DC magnetic field ($\mathbf{H}_0 \parallel \hat{z}$) is swept through the FMR condition of the \Py (Permalloy, Py) layer. We acquire the microwave transmission $S$-parameter $S_{21} \equiv V_\lin{in,2}/V_\lin{out,1}$ where $V_\lin{in(out),1(2)}$ is the voltage input (output) at port 1 (2) of the VNA. Field sweeps were repeated to average the transmission data until an appropriate signal-to-noise ratio was obtained. 

Typically, VNA-FMR measurements focus on the resonance field and linewidth. Our method additionally makes use of the signal magnitude and phase in order to directly probe the AC charge currents produced by iSOT. Previous studies of AC charge currents in spin pumping experiments have relied on intricate experimental setups or techniques that suppress or are insensitive to spurious background signals \cite{weiler_phase-sensitive_2014, wei_spin_2014, hahn_detection_2013}. Our technique remains sensitive to currents induced by the Faraday effect, but is able to separate them from spin-charge conversion currents through the combination of phase-sensitive analysis and comparison to control samples in which the heavy metal NM (here, Pt) is substituted with a Cu layer of nominally negligible intrinsic spin-orbit effects. Furthermore, because the CPW is inductively coupled to the sample, no electrical connections need to be made directly to the FM/NM sample. 

The sample adds a complex inductance $L$ in series with the impedance of the bare CPW, $Z_0$ (here, \SI{50}{\ohm}). The change in microwave transmission $\Delta S_{21}$ is therefore that of a simple voltage divider \cite{silva_characterization_2016}:

\begin{equation}
\Delta S_{21} = -\frac{1}{2}\left(\frac{i \omega L}{Z_0 + i \omega L}\right) \approx \frac{-i \omega L}{2 Z_0}
\label{eq:VoltageDivider}
\end{equation} 

\noindent for $Z_0 >> \omega L$, where $\omega$ is the microwave frequency. The factor of $1/2$ is needed because the port 2 voltage measurement is between the CPW signal and ground (and not between port 2 and port 1). 

\subsection{Inductance Derivations}
\label{sec:IndDerive}

In order to extract values for the SOT effects from the measured $\Delta S_{21}$, we derive expressions for each contribution to $L$. 

\subsubsection{Inductance due to dipole field of dynamic magnetization}
\label{sec:L0}

To derive the inductance due to AC dipolar fields produced by the precessing FM magnetization, we follow Ref. \citenum{silva_characterization_2016}.

\begin{widetext}
\begin{eqnarray}
 {{L}_{0}} &=& \frac{{{\mu }_{0}}\ell }{{{W}_{\text{wg}}}{{d}_{\text{FM}}}{{I}^{2}}}\left[ \int\limits_{-\infty }^{+\infty }{dy\int\limits_{{{d}_{\text{wg}}}}^{{{d}_{\text{FM}}}+{{d}_{\text{wg}}}}{dz\left[ \mathbf{q}\left( y,z \right)\cdot {{\chi}}\left(\omega,{{H}_{0}}\right)\cdot\mathbf{h}_1\left(y,z,I\right)\right]}}\right] \nonumber \\
 && ~~~~~~~~~~~~~~ *\left[\int\limits_{-\infty}^{+\infty}{dy\int\limits_{{{d}_{\text{wg}}}}^{{{d}_{\text{FM}}}+{{d}_{\text{wg}}}}{dz\left[\mathbf{q}\left(y,z\right)\cdot\mathbf{h}_1\left(y,z,I\right)\right]}}\right] \nonumber \\
 &\cong& \frac{{{\mu}_{0}}\ell}{{{W}_{\text{wg}}}{{d}_{\text{FM}}}{{I}^{2}}}{{\chi}_{yy}}\left(\omega,{{H}_{0}}\right)h_{y}^{2}\left(I,z\right)d_{\text{FM}}^{2}W_{\text{wg}}^{2} \nonumber \\
 &\cong& \frac{{{\mu}_{0}}\ell}{{{W}_{\text{wg}}}{{d}_{\text{FM}}}{{I}^{2}}}{{\chi}_{yy}}\left(\omega,{{H}_{0}}\right){{\left(\frac{I}{2{{W}_{\text{wg}}}}\eta\left(z,{{W}_{\text{wg}}}\right)\right)}^{2}}d_{\text{FM}}^{2}W_{\text{wg}}^{2} \nonumber \\
 &=& \frac{{{\mu}_{0}}\ell{{d}_{\text{FM}}}}{4{{W}_{\text{wg}}}}{{\chi}_{yy}}\left(\omega,{{H}_{0}}\right){{\eta}^{2}}\left(z,{{W}_{\text{wg}}}\right) \label{eq:L0}
\end{eqnarray}
\end{widetext}

\noindent where $\mu_0$ is the vacuum permeability, $l$ the sample length, $d_\lin{FM}$ the FM thickness, $W_\lin{wg}$ the width of the CPW signal line (here, \SI{100}{\micro\meter}), and ${\chi}_{yy}(\omega)$ the frequency-dependent magnetic susceptibility. $\eta(z,W_\lin{wg}) \equiv (2/\pi)\arctan(W_\lin{wg}/2z)$ is the spacing loss, ranging from 0 to 1, due to a finite distance $z$ between sample and waveguide. We have assumed the coordinate system described in Fig. \ref{fig:CPWandCoords} ($\hat{x}$ along the CPW signal propagation direction, $\hat{z}$ along the CPW and sample normal). The function $\mathbf{q}(y,z)$ describes the normalized spatial amplitude of the FMR mode. For the uniform mode, $\mathbf{q}(y,z) = 1$ over the entire sample. The first set of integrals in brackets captures the integrated amplitude of the mode as excited by the driving microwave field $\mathbf{h}_1 = h_y \hat{y}$, while the second describes the inductive pickup sensitivity of the CPW. The first approximation assumes uniform microwave field over the sample dimensions. The second approximation utilizes the Karlqvist equation \cite{mallinson_foundations_2012} to approximate the microwave field as $h_y(I,z) \cong I/(2 W_\lin{wg}) \eta(z,W_\lin{wg})$. 


\subsubsection{Inductance due to AC current flow in NM}
\label{sec:L_NM}

Following Rosa \cite{rosa_self_1908}, we model the sample and CPW as two thin current-carrying sheets of width $w = W_\lin{wg}$, separation $z$, and length $l$, so that the mutual inductance is given by

\begin{equation}
L_{12} = \frac{\mu_0}{4 \pi} 2l \left[ \ln\left(\frac{2l}{R}\right) - 1\right]
\label{eq:L12}
\end{equation}

\noindent where $R$ is defined as


\begin{multline}
R \equiv \sqrt{w^2 + z^2} \left( \frac{z}{\sqrt{w^2 + z^2}} \right)^{\left(\frac{z}{w}\right)^2} \\
*\text{exp}\left(\frac{2z}{w} \arctan\left(\frac{w}{z}\right) - \frac{3}{2}\right)
\end{multline}


Viewing the sample-CPW system as a voltage transformer (two mutually-coupled inductors), the voltage induced in the CPW due to current $I_\lin{NM}$ in the NM and the mutual inductance $L_{12}$ is given by $V = -L_{12} (\partial I_\lin{NM}/\partial t)$. If instead we consider the system to be a single lumped-element inductor, the voltage due to the self-inductance contributed by the sample $L_\lin{NM}$ and applied current $I_\lin{CPW}$ is $V = L_\lin{NM} (\partial I_\lin{CPW}/\partial t)$. Therefore, we can calculate $L_\lin{NM}$ as 

\begin{equation}
L_\lin{NM}= -L_{12} \frac{I_\lin{NM}}{I_\lin{CPW}}
\label{eq:LNM_L12}
\end{equation} 

The charge current we are interested in is that driven by the magnetization dynamics of the FM layer, and given by the off-diagonal term of Eq. \ref{eq:OnsagerMatrix}:

\begin{eqnarray}
I_\lin{NM} &=& \hat{x} \cdot \left[ \int\limits_{-d_\lin{NM}}^{+d_\lin{FM}} \mathbf{J}(z) dz \right] W_\lin{wg} \nonumber \\
&=& \hat{x} \cdot \left[ \hat{z} \times (-\sigmaF + \sigmaE - \sigmaO [\hat{m} \times]) \partial_t \hat{m} \right] \nonumber \\
&& *\sgn(\hat{z}\cdot\hat{n}) \left(\frac{\hbar}{2e}\right) W_\lin{wg}
\label{eq:Ic}
\end{eqnarray}

Assuming a linear solution to the Landau-Lifshitz-Gilbert equation of motion for the magnetization, we write a simple relation between the dynamic component of the magnetization $\mathbf{m}$ and microwave field $\mathbf{h}_1$.

\begin{equation}
\partial_t \hat{m} = i \omega \frac{{\chi}}{M_\lin{s}} \mathbf{h}_1
\label{eq:dtm_approx}
\end{equation}

\noindent To convert the vector cross products in Eq. \ref{eq:Ic} to the complex plane, we use ${\chi}$ in the frequency domain \cite{schneider_spin_2007}: 

\begin{widetext}
\begin{equation}
\chi = \dfrac{\gamma \mu_0 \Ms}{\omega_\lin{res}^2 - \omega^2 + i \omega \Delta \omega}
\begin{bmatrix}
\left(1 + \alpha^2\right) \omega_y - i \alpha \omega && i \omega \\
-i \omega && \left(1 + \alpha^2\right)\omega_x - i \alpha \omega
\end{bmatrix}
\label{eq:chi_freq}
\end{equation}
\end{widetext}

\noindent where $\omega_{x,y} \equiv \gamma \mu_0 H_{x,y}$, $H_{x,y}$ is the stiffness field in the $x$ or $y$ direction (including external, anisotropy, and demagnetizing fields), $\omega_\lin{res} \equiv \sqrt{\omega_x \omega_y}$, and $\Delta \omega \equiv \alpha (\omega_x + \omega_y)$. For compactness in the following derivation, we utilize the tensor components of the susceptibility as defined in Eq. \ref{eq:linearSusc}.

Eq. \ref{eq:Ic} has even terms along $\hat{z} \times \partial_t \hat{m}$ and odd terms along $\hat{z} \times (\hat{m} \times \partial_t \hat{m})$. Using Eq. \ref{eq:dtm_approx} for $\partial_t \hat{m}$, we can work out these cross products assuming $\hat{m} \parallel \hat{z}$ (small-angle FMR excitation). The vector components of the even terms are given by:

\begin{eqnarray}
\hat{z} \times \partial_t \hat{m} &=& \hat{z} \times 
\left(
\begin{bmatrix}
\chi_{xx} && \chi_{xy} \\
\chi_{yx} && \chi_{yy}
\end{bmatrix}
\begin{bmatrix}
0 \\
h_y
\end{bmatrix}
\right) \left(\frac{i \omega}{\Ms}\right) \nonumber \\
&=& \hat{z} \times \left( \chi_{xy} h_y \hat{x} + \chi_{yy} h_y \hat{y} \right) \left(\frac{i \omega}{\Ms}\right) \nonumber \\
&=& (-\chi_{yy} h_y \hat{x} + \chi_{xy} h_y \hat{y}) \left(\frac{i \omega}{\Ms}\right) \label{eq:INM_even}
\end{eqnarray}

Similarly, we find for the odd terms:
\begin{eqnarray}
\hat{z} \times (\hat{m} \times \partial_t \hat{m}) &=& \hat{z} \times (-\chi_{yy} h_y \hat{x} + \chi_{xy} h_y \hat{y}) \left(\frac{i \omega}{\Ms}\right) \nonumber \\
&=& (-\chi_{xy} h_y \hat{x} - \chi_{yy} h_y \hat{y}) \left(\frac{i \omega}{\Ms}\right) \label{eq:INM_odd}
\end{eqnarray}

\noindent Noting from Eq. \ref{eq:chi_freq} that $\chi_{xy} = i \chi_{yy}$ (ignoring terms of order $\alpha$ or $\alpha^2$, and working near resonance such that $\omega_x = \omega$), the vector relationships of Eq. \ref{eq:INM_even} and \ref{eq:INM_odd} are substituted into Eq. \ref{eq:Ic}. After evaluating the $\hat{x}$ projection as prescribed by Eq. \ref{eq:Ic} and grouping even and odd terms, we find:

\begin{equation}
I_\lin{NM} =  \left[(\sigmaF - \sigmaE) + i\sigmaO \right]\sgn(\hat{z}\cdot\hat{n}) \dfrac{i \chi_{yy} h_y}{M_\lin{s}} \left(\frac{\hbar \omega}{2e}\right) W_\lin{wg}
\label{eq:I_NM}
\end{equation}

\noindent from which we define $\tilde{\sigma}_\lin{NM} = (\sigmaF - \sigmaE) + i\sigmaO$. On resonance, $\chi_{yy} = -i \gamma \mu_0 \Ms/(2 \alpha_\lin{eff} \omega_\lin{res})$, such that Eq. \ref{eq:I_NM} produces the current phases depicted in Fig. \ref{fig:CPWandCoords}. 

Finally, using the Karlqvist equation \cite{mallinson_foundations_2012}, we approximate the field of the CPW. With these substitutions into Eq. \ref{eq:LNM_L12}, we arrive at the final result for the inductance due to all AC currents in the NM:

\begin{multline}
L_\lin{NM} = \sgn(\hat{z}\cdot\hat{n}) L_{12}(z,W_\lin{wg},l) \eta(z,W_\lin{wg}) \\
*\frac{\hbar\omega}{4 M_\lin{s} e} i {\chi}_{yy}(\omega, H_0) \tilde{\sigma}_\lin{NM}
\label{eq:L_NM}
\end{multline}

The different frequency dependencies of $L_0$ and $L_\lin{NM}$ is critical for our analysis. When normalized to $\chi_{yy}(\omega,H_0)$, $L_0$ is a frequency-independent inductance. By contrast, $L_\lin{NM}$ has an extra factor of $\omega$, reflecting the fact that both Faraday and SOT effects are driven by $\partial_t \hat{m}$, rather than by $\mathbf{m}(t)$ itself. 

Careful attention needs to be paid to the signal phase in order to properly add the inductive effects of $L_0$ and $L_\lin{NM}$. As discussed in detail in the SI Sec. \ref{sec:SigPhase}, it is the current phase in the CPW that determines the propagating signal phase. Using the excitation current in the CPW as the phase reference, we work out the phase of the induced currents due to the perturbative inductance of the sample-on-CPW, and find that the inductances add according to $L = L_0 - i L_\lin{NM}$.

After normalizing by the fitted susceptibility $\tilde{L} \equiv L/{\chi}_{yy}(\omega,H_0)$, the real and imaginary normalized inductance amplitudes are given by:

\begin{eqnarray}
\mbox{Re}(\tilde{L}) &=& \frac{\mu_0 l}{4}\left[ \frac{d_\lin{FM}}{W_\lin{wg}} \eta^2(z,W_\lin{wg}) -  \sgn(\hat{z}\cdot\hat{n}) \eta(z,W_\lin{wg}) \right. \nonumber \\
&& \left. * \frac{L_{12}(z,W_\lin{wg},l)}{\mu_0 l M_\lin{s}} \frac{\hbar \omega}{e}(\sigmaF - \sigmaE) \right] \label{eq:ReL}\\
\mbox{Im}(\tilde{L}) &=& -\frac{\mu_0 l}{4}\bigg[\sgn(\hat{z} \cdot \hat{n}) \eta(z,W_\lin{wg}) \nonumber \\
&&  ~~~~ * \frac{L_{12}(z,W_\lin{wg},l)}{\mu_0 l M_\lin{s}} \frac{\hbar \omega}{e}\sigmaO  \bigg] \label{eq:ImL}
\end{eqnarray}

\noindent Note that when the stacking order of FM and NM is reversed, so is the sign of the SOT and Faraday currents (and therefore their inductance contributions). 

\subsubsection{Magnetization dynamics driven by forward SOT}
From the transformer analogy developed above and discussed in SI Sec. \ref{sec:SigPhase}, we see that ``image currents" are produced in the CPW when currents flow in the conducting sample. Reciprocity requires that the excitation currents in the CPW drive image currents in the sample. This current will produce Amperian torque and forward SOT effects according to Eq. \ref{eq:forwardSOT}, exciting additional magnetization dynamics which are then picked up by the CPW. This series of transduction effects is fully reciprocal with the Faraday and iSOT sequence described above. In the first case, a drive current in the CPW excites magnetization dynamics (via the coupling factor, $\eta(z, W_\lin{wg})$). Those magnetization dynamics drive charge current in the NM via $\tilde{\sigma}_\lin{NM}$. Finally, these charge currents couple into the CPW via the mutual inductance $L_{12}(z, W_\lin{wg},l)$. In the second case, the order is simply reversed: the CPW currents create image currents in the NM (via $L_{12}(z, W_\lin{wg},l)$), which drive magnetization dynamics (via $\tilde{\sigma}_\lin{NM}$), which are picked up by the CPW (via $\eta(z, W_\lin{wg})$). It can be shown that the induced signal due to forward Amperian or SOT-driven magnetization dynamics add together in-phase with their inverse counterparts, increasing the inductive response from each contribution by a factor of 2. The inductance in Eq. \ref{eq:L_NM} (and hence \ref{eq:ReL} and \ref{eq:ImL}) is therefore too small by a factor of 2. Therefore, in the below calculation of $\tilde{\sigma}_\lin{NM}$ based on measured values of $\tilde{L}_\lin{NM}$, we include this factor.

\subsection{Background Correction}
\label{sec:BGCorr}
To make use of the phase and amplitude information in the VNA-FMR spectra, we first fit the raw spectra to:

\begin{equation}
S_{21}(\omega, H_0) = A e^{i\phi} {\chi}_{yy}(\omega, H_0) + {C}_0 + {C}_1 H_0
\label{eq:S21fit} 
\end{equation}

\noindent where $A$ is the signal amplitude, $\phi$ is the raw phase (inclusive of signal line delay), and ${C}_0$ and ${C}_1$ are complex offset and slope corrections to the background. Utilizing the information in this complex background is key to our data processing method. The background-corrected signal can be plotted from the measured values of $S_{21}$ as:

\begin{equation}
\Delta S_{21}(\omega, H_0) = \frac{S_{21}(\omega, H_0) - (C_0 + C_1 H_0)}{C_0 + C_1 H_0} 
\label{eq:BGcorr_plot}
\end{equation}







\begin{figure}
	\centering
	\includegraphics[width=\linewidth]{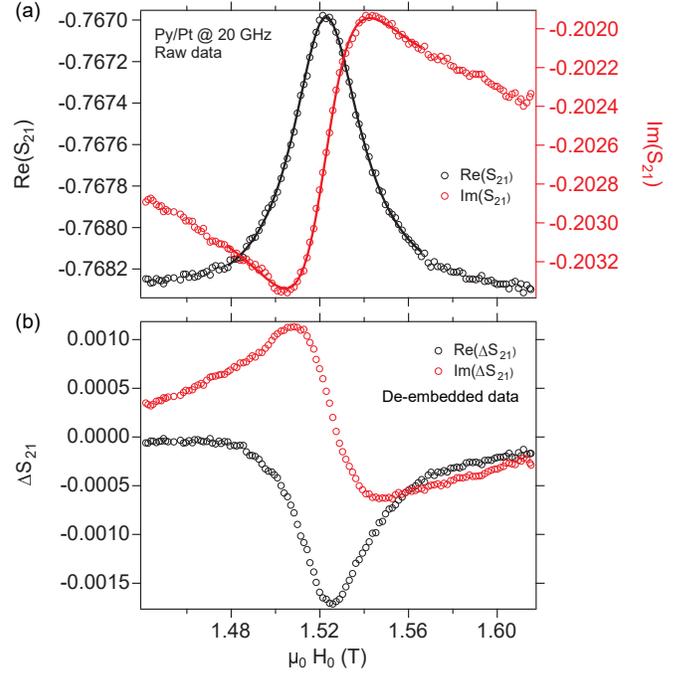}
	\caption{Example $S_{21}$ spectrum, acquired at f = \SI{20.0}{\giga\hertz}. (a) Raw data, with fits. Note the different background offsets of the Re and Im data (left and right axes). (b) De-embedded $\Delta S_{21}$ signal.}
	\label{fig:S21_BGcorr}
\end{figure}

\noindent This corrects the signal phase for the finite length of the signal line between the VNA source and receiver ports and the sample, effectively placing the ports at the sample position. Additionally, it normalizes the signal amplitude by the frequency-dependent losses due to the complete microwave circuit (cables + CPW + sample). In Fig. \ref{fig:S21_BGcorr}(a) and (b), we plot the raw and de-embedded data, respectively. The large complex offset on top of which the resonance signal is superimposed in (a) represents $C_0$ and $C_1$. 

Comparison of Eqs. \ref{eq:S21fit} and \ref{eq:BGcorr_plot} shows that the change in microwave transmission can be written as:

\begin{equation}
\Delta S_{21}(\omega, H_0) = \frac{A e^{i\phi}}{{C}_0 + {C}_1 H_0} {\chi}_{yy}(\omega, H_0)
\label{eq:BGcorr_calc}
\end{equation}

\noindent Using this form for the background-corrected $\Delta S_{21}$, the inductance amplitude $\tilde{L}(f)$ is calculated as $[\Delta S_{21}/{\chi}_{yy}(\omega, H_0)] [i 2 Z_0/(2\pi f)]$. When $\tilde{L}$ is plotted vs. frequency as in Fig. \ref{fig:ReImL}, we note that there can be a small phase error that causes $\mbox{Im}(\tilde{L})(f\rightarrow0) \neq 0$. The correction for this phase error is discussed in SI Sec. \ref{sec:AnPhase}.


\subsection{Calculation of $\tilde{\sigma}_\lin{NM}$ from measured $L$}
Using the results for $\mbox{Re}(\tilde{L})$ and $\mbox{Im}(\tilde{L})$ (Eqs. \ref{eq:ReL} and \ref{eq:ImL}), we can isolate the $\tilde{\sigma}_\lin{NM}$ contribution as follows. First, the slope of $\tilde{L}$ is used to isolate the contribution of $\tilde{L}_\lin{NM}$:

\begin{multline}
\frac{d\tilde{L}}{df} = - \frac{1}{2}\sgn(\hat{z}\cdot\hat{n}) \eta(z,W_\lin{wg}) \frac{L_{12}(z,W_\lin{wg},l)}{M_\lin{s}}\\
*\frac{h}{e} \left[(\sigmaF - \sigmaE) + i \sigmaO \right]
\end{multline}

\noindent We normalize $d\tilde{L}/df$ by $\tilde{L}_0$ in order to remove any residual differences in sample-CPW coupling from sample to sample. Variation in $\tilde{L}_0$ (e.g., as seen in Fig. \ref{fig:ReImL}) can be caused by sample-to-sample variations in magnetization, including dead layer effects at the various FM/NM interfaces, as well as measurement-to-measurement variations in the sample-waveguide spacing, which could be affected by small dust particles in the measurement environment. Finally, we solve for the effective conductivity. 

\begin{multline}
\left[(\sigmaF - \sigmaE) + i \sigmaO \right] = -\sgn(\hat{z}\cdot\hat{n}) \left(\frac{\dfrac{d\tilde{L}}{df}}{2 \tilde{L}_0}\right) \\
*\frac{\mu_0 l}{L_{12}(z,W_\lin{wg},l)} \frac{M_\lin{s} d_\lin{FM}}{W_\lin{wg}} \eta(z,W_\lin{wg})\frac{e}{h}
\label{eq:sigma_from_L}
\end{multline}

\subsection{Analysis Protocol}
\label{sec:Protocol}
Our quantitative VNA-FMR analysis protocol is summarized below \cite{Note1}.

\begin{enumerate}
\setlength\itemsep{0em}
\item Complex VNA-FMR data is collected and fit with Eq. \ref{eq:S21fit}.
\item $\Delta S_{21}$ is calculated with Eq. \ref{eq:BGcorr_calc} to de-embed the sample contribution to the inductance.
\item $\Delta S_{21}$ is converted to sample inductance $L$ using Eq. \ref{eq:VoltageDivider}.
\item $L$ is normalized by magnetic susceptibility $\chi_{yy}$, yielding the complex inductance amplitude given by Eqs. \ref{eq:ReL} and \ref{eq:ImL} ($\mbox{Re}(\tilde{L})$ and $\mbox{Im}(\tilde{L})$).
\item The phase error of $\tilde{L}$ is corrected as described in SI Sec. \ref{sec:AnPhase}.
\item Linear fits of $\tilde{L}(\omega)$ (using Eqs. \ref{eq:ReL} and \ref{eq:ImL}) are used to extract $\tilde{L}_0$ and $\tilde{L}_\lin{NM}(\omega)$.
\item The effective conductivities $\sigmaO$ and $(\sigmaF - \sigmaE)$ are obtained from $(\partial \tilde{L}/\partial f)/\tilde{L}_0$ according to Eq. \ref{eq:sigma_from_L}.
\end{enumerate}

\section{Data and Analysis}
\label{sec:Data}
\begin{figure}
	\centering
	\includegraphics[width=\linewidth]{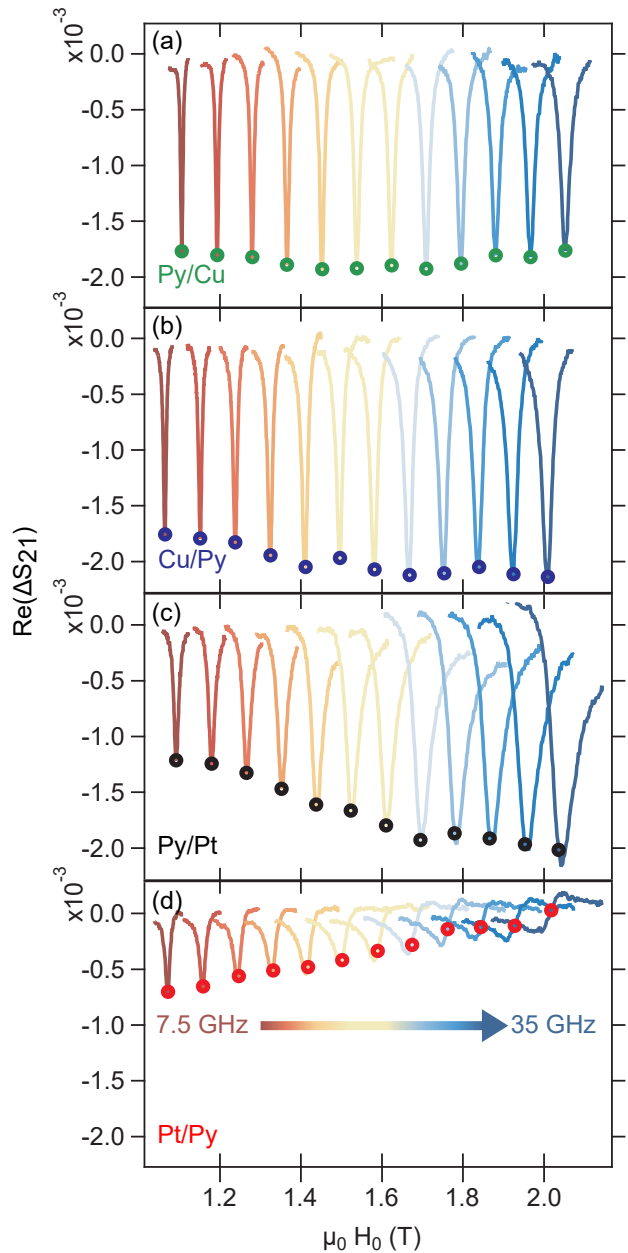}
	\caption{FMR spectra for FM/NM bilayers. $\mbox{Re}(\Delta S_{21})$ at several excitation frequencies for different samples: (a) Py/Cu, (b) Cu/Py, (c) Py/Pt, and (d) Pt/Py. The change in lineshape and amplitude for Py/Pt and Pt/Py clearly shows the presence of frequency-dependent inductive terms not present in the Py/Cu and Cu/Py control samples. The colored circles indicate the value of $\mbox{Re}(\Delta S_{21})\propto \mbox{Re}(L)$ when $H_0$ satisfies the out-of-plane FMR condition.}
	\label{fig:S21_evolution}
\end{figure}

To demonstrate the quantitative VNA-FMR technique, we measured FMR in metallic stacks consisting of substrate\slash{}Ta(1.5)\slash{}Py(3.5)\slash{}NM\slash{}Ta(3) and inverted stacks of substrate\slash{}Ta(1.5)\slash{}NM\slash{}Py(3.5)\slash{}Ta(3) (where the numbers in parentheses indicate thickness in nanometers). We focus on a Pt(6) NM layer due to its large intrinsic SOC, and use Cu(3.3) as a control material with nominally negligible SOC \cite{saitoh_conversion_2006, niimi_extrinsic_2011, sinova_spin_2015}. We collected room-temperature FMR spectra as a function of out-of-plane external magnetic field $H_0$ with microwave frequencies from \SI{5}{\giga\hertz} to \SI{35}{\giga\hertz} and VNA output power of 0 dBm. Exemplary $\mbox{Re}(\Delta S_{21})$ spectra are shown in Fig. \ref{fig:S21_evolution}. Each raw spectrum has been normalized by the complex signal background (see Sec. \ref{sec:BGCorr}). In the following discussion, we use a notation for the bilayers which indicates the sample growth order as read from left-to-right. For example, Py/Pt indicates Py is first deposited onto the substrate, followed by Pt.

\begin{figure}
	\centering
	\includegraphics[width=\linewidth]{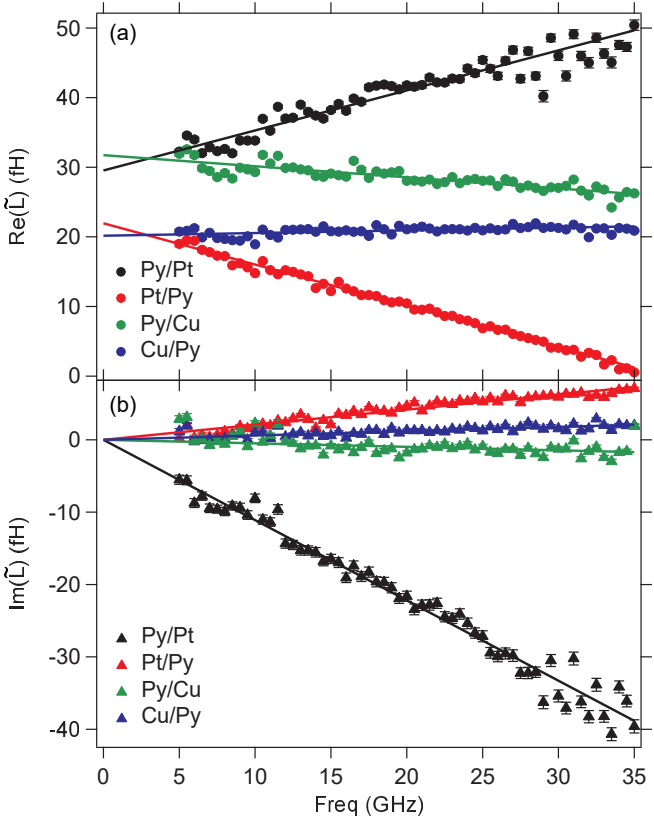}
	\caption{Frequency dependence of real and imaginary inductances extracted from $S_{21}$ spectra (symbols) and fits to Eqs. \ref{eq:ReL} and \ref{eq:ImL} (lines). (a) $\mbox{Re}(\tilde{L})$ for all samples with $l=$ \SI{8}{\milli\meter}. Zero-frequency intercept indicates the dipolar inductive coupling, while the linear slope reflects $(\sigmaF - \sigmaE)$. (b) $\mbox{Im}(\tilde{L})$ for all samples, as a function of frequency, where the linear slope is governed by $\sigmaO$.}
	\label{fig:ReImL}
\end{figure}

Both Py/Cu and Cu/Py samples exhibit a mostly real normalized inductance amplitude (symmetric Lorentzian dip for $\mbox{Re}(\Delta S_{21})$ in Fig. \ref{fig:S21_evolution}(a) and (b)) with a magnitude largely independent of frequency, in accordance with $\tilde{L}_\lin{NM} \approx 0$. That is, the signal is dominated by the dipolar inductance. In contrast, the lineshape and magnitude of the Py/Pt and Pt/Py data in Fig. \ref{fig:S21_evolution}(c) and (d) exhibit a clear frequency dependence as expected for $\tilde{L}_\lin{NM} \neq 0$. In particular, the data for Py/Pt indicate that $\tilde{L}_\lin{NM}$ adds constructively with $L_0$, such that $\mbox{Re}(\tilde{L})$ increases with increasing $f$. The Pt/Py inductance evolves in an opposite sense due to the stack inversion, leading to a decrease and eventual compensation of $\mbox{Re}(\tilde{L})$ at high $f$. The increasingly antisymmetric lineshape for both Py/Pt and Pt/Py reveals that the magnitude of $\mbox{Im}(\tilde{L})$ also increases with frequency, with a sign given by the stacking order. 

By normalizing the spectra in Fig. \ref{fig:S21_evolution} to the magnetic susceptibility $\chi(\omega,H_0)$ defined in Eq. \ref{eq:chi}, we extract the complex inductance amplitude $\tilde{L}$. $\mbox{Re}(\tilde{L})$ and $\mbox{Im}(\tilde{L})$ are shown in Fig. \ref{fig:ReImL} for all investigated bilayers with a length $l$ of \SI{8}{\milli\meter}. As shown in Eqs. \ref{eq:ReL} and \ref{eq:ImL}, $\mbox{Re}(\tilde{L})$ provides information about the dipolar inductance ($\tilde{L}_0$, zero-frequency intercept), and $-(\sigmaF - \sigmaE)$ (slope). Similarly, the slope of $\mbox{Im}(\tilde{L})$ reflects $-\sigmaO$. Immediately evident is the reversal of the slope for Py/Pt compared to Pt/Py, which is captured by the $\sgn$ function (where $\hat{n}$ is the FM/NM interface normal, pointing into the FM, and $\hat{z}$ is defined by the coordinate system in Fig. \ref{fig:CPWandCoords}). This sign-reversal is consistent with the phenomenology expected for interface-symmetry sensitive effects, e.g., combined spin pumping and iSHE, as well as iREE. There is also a marked difference in the slope magnitude for Py/Pt and Pt/Py in panel (b), the implications of which are discussed below. 

\begin{figure}
	\centering
	\includegraphics[width=\linewidth]{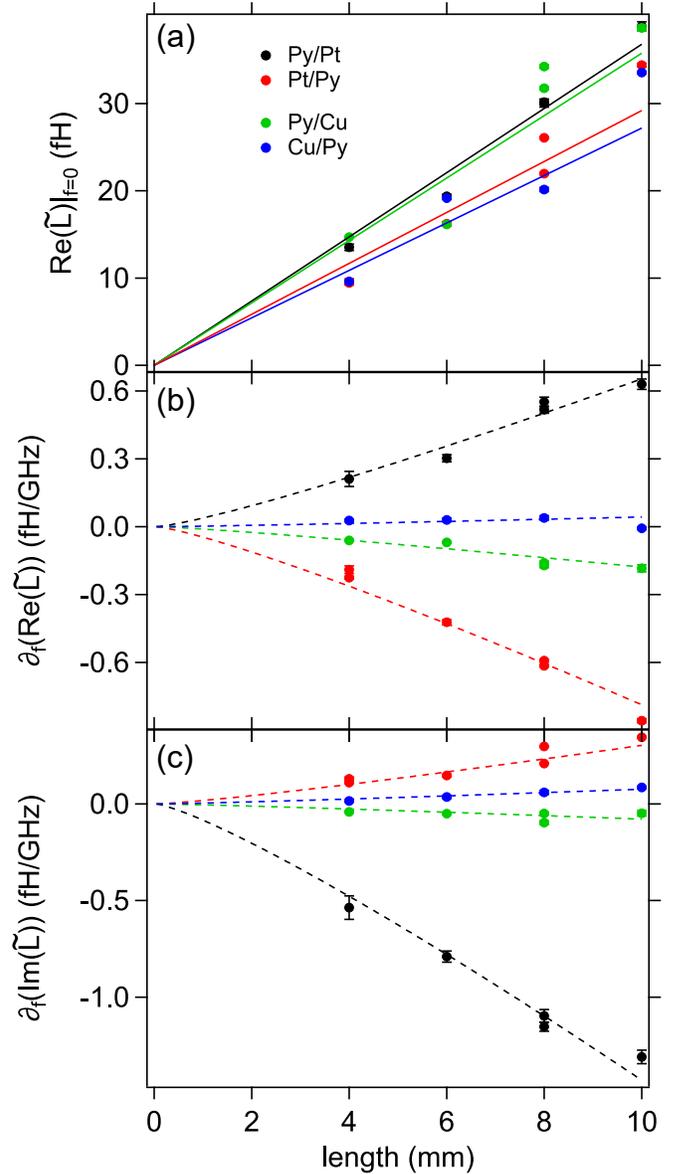}
	\caption{$\tilde{L}(f=0)$ and $\partial \tilde{L}/\partial f$ extracted from data as in Fig. \ref{fig:ReImL} vs. sample length for all samples. (a) Dipolar inductive coupling $\tilde{L}_0$. (b) From $\partial[\mbox{Re}(\tilde{L})]/\partial f$, we extract $(\sigmaF - \sigmaE)$. (c) From $\partial [\mbox{Im}(\tilde{L})]/\partial f$, we extract $\sigmaO$. Dashed lines are guides based on Eqs. M8 and M9 with values of $\sigmaO$ and $(\sigmaF - \sigmaE)$ calculated as described in the Methods. Several measurements were repeated  to demonstrate reproducibility.}
\label{fig:Z_lengthDep}
\end{figure}

Each of the inductance terms has some dependence on sample length: linear for the dipolar contribution, and slightly non-linear for the inductances due to charge flow in the NM (see Eqs. \ref{eq:L0} and \ref{eq:L12}). We therefore repeated the measurements shown in Fig. \ref{fig:ReImL} for a variety of sample lengths from 4 to \SI{10}{\milli\meter}. Fig. \ref{fig:Z_lengthDep} shows the measured inductance terms $\tilde{L}_0$, $\partial \mbox{Re}(\tilde{L})/\partial f$ (intercept and slope of curves in Fig. \ref{fig:ReImL}(a)), and $\partial \mbox{Im}(\tilde{L})/\partial f$ (slope of curves in Fig. \ref{fig:ReImL}(b)) as a function of sample length. Following normalization by its corresonding $\tilde{L}_0$, each data point in Fig. \ref{fig:Z_lengthDep}(b) provides a value of $(\sigmaF - \sigmaE)$ (see Eq. \ref{eq:sigma_from_L}). Similarly, data points in panel (c) provide values of $\sigmaO$. These values are averaged to provide a single $(\sigmaF - \sigmaE)$ and $\sigmaO$ for each sample type. Results are summarized in Table \ref{tab:SpinChargeParams}. The dashed lines in Fig. \ref{fig:Z_lengthDep}(b) and (c) are calculated curves based on these average values and the length dependence of $\tilde{L}$.

Because $\sigmaE$ and $\sigmaF$ have the same phase and frequency dependence, we use control samples where we replace the Pt with Cu, wherein it is generally accepted that both the SHE for Cu and the REE at the Py/Cu interface are negligible \cite{saitoh_conversion_2006, niimi_extrinsic_2011, sinova_spin_2015}. Furthermore, the Cu thickness is chosen so that it exhibits the same sheet resistance as the Pt layer, so that the two samples have identical $\sigmaF$ (see Eq. \ref{eq:sigmaF}). Subtraction of the time-reversal-even conductivity for the Py/Cu control samples from the time-reversal-even conductivity for the Py/Pt samples therefore isolates $\sigmaE$ specifically for the Py/Pt interface. Likewise, any damping-like contributions to $\sigmaO$ due to the Ta seed layer should also be removed by subtraction of the Py/Cu inductance data.

Additional data collected for varied NM thickness (to be presented in a future publication) indicates that the charge currents produced by iSOT effects experience a shunting effect, whereby some fraction of the interfacial charge current flows back through the sample thickness, reducing the inductive signal. This can be modeled as a current divider with some of the iSOT-generated current coupling to the \SI{50}{\ohm} CPW via image currents, and the remainder shunted by the sheet conductance of the sample. Final values of the extracted conductivities reported in Table \ref{tab:SpinChargeParams} have been corrected to account for current shunting (see SI Sec. \ref{sec:ShuntCorr} for more details). Comparison of the shunt-corrected SOT conductivities makes evident that the field-like charge currents are comparable to those due to damping-like spin-charge conversion processes. 

We can compare our measured values of $\sigmaE$ and $\sigmaO$ to measurements made by other groups using different techniques. Garello, et al.\cite{garello_symmetry_2013} use the harmonic Hall technique and Miron, et al. \cite{mihai_miron_current-driven_2010} investigate domain wall nucleation to quantify the spin-orbit torque exerted on Co sandwiched between Pt and AlO$_x$. Converting their measured values of field-like SOT field per unit current density to our metric $\sigmaE$, they find \SI{1.1e6}{\ohm^{-1}\meter^{-1}} and \SI{1.9e7}{\ohm^{-1}\meter^{-1}}. Nguyen, et al. \cite{nguyen_spin_2016} find a similar value of $\approx$ \SI{1.3e6}{\ohm^{-1}\meter^{-1}} for a Pt/Co bilayer using harmonic Hall methods. The Garello and Nguyen results are within an order of magnitude of our findings (\SI{-1.48(7)e5}{\ohm^{-1}\meter^{-1}} for Pt/Py and \SI{-1.8(2)e5}{\ohm^{-1}\meter^{-1}} for Py/Pt).

Garello and Nguyen also report damping-like values for their effective SOT fields. Converted to $\sigmaO$, they find \SI{5.8e5}{\ohm^{-1}\meter^{-1}} and $\approx$\SI{2.9e5}{\ohm^{-1}\meter^{-1}}, respectively, which are again within an order of magnitude of our values: \SI{2.4(3)e5}{\ohm^{-1}\meter^{-1}} (Py/Pt) and \SI{0.6(2)e5}{\ohm^{-1}\meter^{-1}} (Pt/Py).




\begin{table*}
\begin{centering}
\begin{tabular}{ l S S S S }
\toprule
Sample & {$(\sigmaF - \sigmaE)_\lin{meas}$} & {$(\sigmaO)_\lin{meas}$} & {$(\sigmaE)_\lin{corr}$} & {$(\sigmaO)_\lin{corr}$} \\
\hline
Py/Pt & -0.45(3) & 1.0(1) & -1.48(7) & 2.4(3)  \\
Pt/Py & -0.69(5) & 0.31(6) & -1.8(2) & 0.6(2)  \\
Py/Cu & 0.143(6) & 0.07(3) & & \\
Cu/Py & 0.04(3) & 0.06(1) & & \\
\botrule
\end{tabular}
\caption{Effective conductivities (in units of $10^5$ \SI{}{\ohm^{-1} \meter^{-1}}) and microscopic spin-charge conversion parameters (Rashba parameter $\alpha_\lin{R}$ and spin Hall angle $\theta_\lin{SH}$). Measured values are calculated from measured inductances (Fig. \ref{fig:Z_lengthDep}). Corrected values are calculated by subtraction of Cu control to remove the Faraday contribution (in the case of $\sigma_\lin{e}$) and any contribution from the Ta interfaces, followed by application of the shunting correction (see SI Sec. \ref{sec:ShuntCorr}).}
\label{tab:SpinChargeParams}
\end{centering}
\end{table*}

\section{Discussion}
\label{sec:Discuss}
For comparison to previous measurements and to theory, we can relate the effective conductivities $\sigmaE$ and $\sigmaO$ to microscopic spin-charge conversion parameters under the assumptions that the damping-like iSOT is due to iSHE only, and the field-like iSOT is from iREE only. We also relate the Faraday contribution to the AC charge currents in the NM---that is, $\sigmaF$---to sample properties. 

\subsection{Contributions to effective conductivity, $\tilde{\sigma}_\lin{NM}$}
\label{sec:EffConductivity}

\subsubsection{Effective Faraday conductivity, $\sigmaF$}
To relate the effective Faraday conductivity, $\sigmaF$, to sample parameters, we isolate the Faraday component of the induced charge current from Eq. \ref{eq:inverseSOT}: 

\begin{equation}
\left[\displaystyle\int\limits_{-d_\lin{NM}}^{+d_\lin{FM}} \mathbf{J}^{\lin{F}}(z) dz\right] = -\sgn(\hat{z}\cdot\hat{n}) \left(\frac{\hbar}{2e}\right) \sigmaF(\hat{z}\times \partial_t  \hat{m})
\label{eq:JFar}
\end{equation}

\noindent The charge current is driven by the induced e.m.f., $V_x$, according to:

\begin{eqnarray}
\hat{x} \cdot \left[\displaystyle\int\limits_{-d_\lin{NM}}^{+d_\lin{FM}} \mathbf{J}^{\lin{F}}(z) dz\right] &=& \frac{I_x}{w} \nonumber \\
&=& \frac{V_x}{Z_\lin{eff} l} \label{eq:inducedEMF}
\end{eqnarray}

\noindent The induced e.m.f. is derived from inductive reciprocity \cite{wessel-berg_generalized_1978}

\begin{equation}
V_x = -\frac{\partial \phi}{\partial t} = - \mu_0 M_\lin{s} \int\limits_{V_\lin{FM}} [ \mathbf{h(r)} \cdot \partial_t \hat{m} ] d^3 r 
\label{eq:inductiveRecip}
\end{equation}

\noindent where $\mathbf{h}(\mathbf{r})$ is the magnetic sensitivity function for a current of unit amplitude in the NM layer. We assume this field can be approximated with the Karlqvist equation, and use the results for $\partial_t \hat{m}$ from Sec. \ref{sec:IndDerive}. Subsituting Eq. \ref{eq:inductiveRecip} into Eq. \ref{eq:inducedEMF}, and equating the result with Eq. \ref{eq:JFar} yields the final expression for $\sigmaF$:

\begin{equation}
\sigmaF = \frac{e \mu_0 M_\lin{s} d_\lin{FM}}{\hbar Z_\lin{eff}}
\label{eq:sigmaF}
\end{equation}

\subsubsection{Rashba parameter and $\sigmaE$}

We can relate the even spin-orbit torque conductivity $\sigmaE$ to the Rashba parameter $\alpha_\lin{R}$. We start from the field-like interfacial spin torque per spin $\mathbf{t}_\lin{fl}$ introduced by Kim, et al. (Eq. 12 in Ref. \citenum{kim_chirality_2013}):

\begin{equation}
\mathbf{t}_\lin{fl} = \sgn(\hat{z}\cdot\hat{n})k_\lin{R} v_\lin{s} \left[ \hat{m} \times (\hat{j} \times \hat{z}) \right] \left(\frac{\hbar}{2}\right)
\label{eq:KimSpinTorque}
\end{equation}

\noindent where $k_\lin{R} = 2 \alpha_\lin{R} m_\lin{e}/\hbar^2$ is a wavevector corresponding to the Rashba energy parameter $\alpha_\lin{R}$, $m_\lin{e}$ is the mass of the electron, and $v_\lin{s} = P J_\lin{int} g \mu_\lin{B}/(2 e M_\lin{s})$ is the spin velocity, with charge current density $J_\lin{int}$ at the FM/NM interface at which the Rashba effect is present, spin polarization of the charge current $P$, Land\'e g-factor $g$ , and Bohr magneton $\mu_\lin{B}$. Note that $\mathbf{t}_\lin{fl}/(\hbar/2)$ has units of \SI{}{\hertz}; that is, the same units as $\partial_t \hat{m}$. We can therefore relate Eq. \ref{eq:KimSpinTorque} to the volume-averaged magnetic torque density $\mathbf{T}$ from Eqs. \ref{eq:LLG} and \ref{eq:forwardSOT} through the time rate of change of the magnetization: $\mathbf{t}_\lin{fl} d_\lin{int} \delta(z)/(\hbar/2) = \partial_t \hat{m}$, where we have added $d_\lin{int} \delta(z)$ to account for the interfacial nature of this torque (where $d_\lin{int}$ is an effective thickness of the interface).  

\begin{eqnarray}
\frac{2}{\hbar} \int\limits_0^{d_\lin{FM}} \mathbf{t}_\lin{fl} d_\lin{int} \delta(z) dz &=& -\frac{\gamma}{M_\lin{s}}\int\limits_0^{d_\lin{FM}}\mathbf{T}(z) dz \label{eq:TorqueFL} \\
k_\lin{R} v_\lin{s} \hat{m} \times (\hat{j} \times \hat{z}) d_\lin{int} &=& -\frac{\gamma}{M_\lin{s}} \frac{\hbar}{2e} \sigmaE \hat{m} \times (\hat{z} \times \mathbf{E}) \label{eq:InterfacialTorque}
\end{eqnarray}

\noindent The final line results from subsituting Eq. \ref{eq:KimSpinTorque} and the even SOT term from Eq. \ref{eq:forwardSOT} into Eq. \ref{eq:TorqueFL}. Making the substitutions for $k_\lin{R}$ and $v_\lin{s}$, and using $\mathbf{E} = (J_\lin{int}/\sigma_\lin{int})\hat{j}$ yields: 

\begin{equation}
\alpha_\lin{R} = \dfrac{\hbar^2}{2 m_\lin{e}}\dfrac{\sigmaE}{\sigma_\lin{int}}\dfrac{1}{P d_\lin{int}}
\label{eq:alpha_Rashba}
\end{equation}

\noindent Here, $\sigma_\lin{int}$ is the interfacial conductivity of the FM/NM interface (extracted by measuring resistance vs. Py thickness; see SI Sec. \ref{sec:IntResistivity}) and $P$ is the spin polarization at the FM/NM interface. We use $P = 0.6$ as determined via spin-wave Doppler measurements in Ref. \citenum{zhu_temperature_2010}, and assume $d_\lin{int}$ is one Py lattice constant (\SI{0.354}{\nano\meter}) \cite{haney_current-induced_2013}. We therefore find $\alpha_\lin{R}=$ \SI{-5.8(3)}{\milli\electronvolt.\nano\meter} for the Py/Pt sample, and \SI{-7.5(7)}{\milli\electronvolt.\nano\meter} for Pt/Py. These values are smaller than those measured with angle-resolved photoelectron spectroscopy (ARPES) for the surface state of Au(111) (\SI{33}{\milli\electronvolt.\nano\meter}) \cite{cercellier_interplay_2006}, Bi(111) (\SI{56}{\milli\electronvolt.\nano\meter}) \cite{koroteev_strong_2004}, and Ge(111) (\SI{24}{\milli\electronvolt.\nano\meter}) \cite{yaji_large_2010}, and much smaller than the Bi/Ag(111) interface (\SI{305}{\milli\electronvolt.\nano\meter}) \cite{ast_giant_2007}.





We can also compare our results for the Rashba parameter to a recent theoretical calculation. Kim, Lee, Lee, and Stiles (KLLS) \cite{kim_chirality_2013} have shown that SOT and the Dzyaloshinskii-Moriya interaction (DMI) at a FM/NM interface are both manifestations of an underlying Rashba Hamiltonian, and predict a straightfoward relationship between  the Rashba parameter $\alpha_\lin{R}$, interfacial DMI strength $D_\lin{DMI}^\lin{int}$, and the interfacial field-like SOT per spin $t_\lin{fl}$:

\begin{equation}
\alpha_\lin{R} =\frac{\hbar^2}{2 m_\lin{e}} \left( \frac{D_\lin{DMI}^\lin{int}}{2A} \right) = \frac{\hbar}{m_\lin{e}} \left(\frac{t_\lin{fl}}{v_\lin{s}}\right)
\label{eq:RashbaPredict}
\end{equation}

\noindent where $A$ is the exchange stiffness. 

For the Pt/Py stack, the ratio of interfacial DMI, $D_\lin{DMI}^\lin{int}$, to bulk exchange $A$ was previously measured via a combination of Brillouin light scattering (BLS) and superconducting quantum interference device (SQUID) magnetometry for samples prepared under nearly identical growth conditions, albeit with a stack geometry that was optimized for optical BLS measurements \cite{nembach_linear_2015}. The ratio is a constant value of \SI{-0.25(1)}{\nano\meter^{-1}} over a Py thickness range of 1.3 to \SI{15}{\nano\meter}. As such, this material system is an ideal candidate to test the quantitative prediction of the KLLS theory. Using the experimentally-determined value for $D_\lin{DMI}^\lin{int}/A$ with Eq. \ref{eq:RashbaPredict} predicts a Rashba strength of \SI{-4.8(2)}{\milli\electronvolt.\nano\meter}, which agrees well in sign and magnitude with the result of our iSOT measurement for the Pt/Py sample of the same stacking order, as well as the Py/Pt sample with opposite stacking order. Together, the spin wave spectroscopy and iSOT measurements clarify the role of the Rashba spin-orbit interaction as the underlying physical mechanism for both DMI and field-like SOT in the Py/Pt system.

\subsubsection{Spin Hall angle and $\sigmaO$}
In order to develop intuition for Eq. \ref{eq:inverseSOT} we first derive an approximate relationship between $\sigmaO$ and the spin Hall angle, $\theta_\lin{SH}$, applicable when the NM thickness is much thicker than its spin diffusion length. We assume series resistors $1/G_{\uparrow\downarrow} + 1/G_\lin{ext}$ (interfacial spin-mixing conductance + spin conductance of the NM) in a  voltage divider model for the spin accumulation at the FM/NM interface due to spin pumping

\begin{equation}
\mu_\lin{s}(z=0^+) \hat{s} = \frac{\hbar}{2}\left(\hat{m} \times \frac{\partial \hat{m}}{\partial t}\right)\left( \frac{G_{\uparrow\downarrow}}{G_{\uparrow\downarrow} + G_\lin{ext}} \right)
\label{eq:SpinPumpedAccum}
\end{equation}

\noindent where $\mu_\lin{s}(z=0^+)$ is the spin accumulation at the FM/NM interface. Using the result of Eq. 6 from Ref. \citenum{boone_spin_2013} for the effective one-dimensional spin conductance of a NM (where we have set $G_2^\lin{NM} = 0$ because we are interested in only a FM/NM bilayer, not a FM/NM1/NM2 multilayer):

\begin{equation}
G_\lin{ext} = \frac{\sigma}{2 \lambda_\lin{s}}\tanh\left(\frac{d_\lin{NM}}{\lambda_\lin{s}}\right)
\label{eq:Gext}
\end{equation}

\noindent where $\lambda_\lin{s}$ is the spin diffusion length in the NM. The integrated charge current in the NM layer driven by the resulting spin chemical potential gradient $-\nabla \mu_\lin{s} = \mathbf{Q}_s$ and the inverse spin Hall effect ($\mathbf{J}_c \propto \mathbf{Q}_s \times \hat{s}$) is given by 

\begin{eqnarray}
\int\limits_0^{d_\lin{NM}} \textbf{J}_\lin{c}(z) dz &=& \int\limits_0^{d_\lin{Pt}} \left[ \sigma_\lin{SH} \frac{-\nabla \mu_\lin{s}(z)}{e} \times \hat{s} \right] dz \\
&=& \sigma_\lin{SH} \dfrac{\mu_\lin{s}(z = 0^+)}{e} (-\hat{z} \times \hat{s})\label{eq:intSHcurrent}
\end{eqnarray}

\noindent assuming $d_\lin{NM} >> \lambda_\lin{s}$. The spin Hall conductivity is related to the spin Hall angle via the Pt charge conductivity: $\sigma_\lin{SH} = \theta_\lin{SH} \sigma_\lin{Pt}$. If we combine Eqs. \ref{eq:SpinPumpedAccum}, \ref{eq:Gext}, and \ref{eq:intSHcurrent} and equate the integrated charge current to that from $\sigmaO$ in Eq. \ref{eq:inverseSOT} we arrive at the final result:

\begin{equation}
\sigmaO = \sigma \left\{\theta_\lin{SH} \mbox{Re}\left[ \frac{G_{\uparrow \downarrow}}{\dfrac{\sigma}{2 \lambda_\lin{s}} \tanh\left(\dfrac{d_\lin{NM}}{\lambda_\lin{s}}\right) + G_{\uparrow \downarrow}} \right]\right\} \epsilon
\label{eq:SHAngle}
\end{equation}

\noindent The model also accounts for less-than-unity efficiency $\epsilon$ for spin transmission into the NM (such that $(1-\epsilon)$ is the spin loss fraction, which has been attributed to processes such as spin memory loss \cite{rojas-sanchez_spin_2014} or promixity magnetism \cite{caminale_spin_2016}). 

A more accurate version of Eq. \ref{eq:SHAngle} is obtained by replacing the unitless term in curly brackets with Eq. 11 from Ref. \citenum{haney_current_2013}:

\begin{multline}
\sigmaO = \sigma \left\{\theta_\lin{SH} \frac{(1 - e^{-d_\lin{NM}/\lambda_\lin{s}})^2}{(1 + e^{-2 d_\lin{NM}/\lambda_\lin{s}})}\right. \\
\left.*\frac{|\tilde{G}_{\uparrow\downarrow}|^2 + \mbox{Re}(\tilde{G}_{\uparrow\downarrow})\tanh^2\left(\dfrac{d_\lin{NM}}{\lambda_\lin{s}}\right)}
{|\tilde{G}_{\uparrow\downarrow}|^2 + 2 \mbox{Re}(\tilde{G}_{\uparrow\downarrow})\tanh^2\left(\dfrac{d_\lin{NM}}{\lambda_\lin{s}}\right) + \tanh^4\left(\dfrac{d_\lin{NM}}{\lambda_\lin{s}}\right)}\right\} \epsilon
\label{eq:StilesHaney}
\end{multline}

\noindent where $\tilde{G}_{\uparrow\downarrow} = G_{\uparrow\downarrow} 2 \lambda_\lin{s} \tanh(d_\lin{NM}/\lambda_\lin{s})/\sigma$. This properly accounts for the boundary condition that the spin current goes to zero at the distant surface of the NM. 

Eq. \ref{eq:StilesHaney} can be used to calculate $\theta_\lin{SH}$ if we assume values for $\lambda_\lin{s}$, $G_{\uparrow\downarrow}$, and $\epsilon$. If these parameters are presumed identical for the two stacking orders, we would find spin Hall angles that differ by a factor of 4 depending on whether Pt is deposited on Py, or vice versa. Instead, the large discrepancy in $\sigmaO$ for the two stacking orders suggests differences in the FM/NM interface that affect  $G_{\uparrow\downarrow}$ and $\epsilon$. Given the data presented here, it is possible for us to estimate the efficiency with which spins are pumped into the Pt layer as follows. The total Gilbert damping $\alpha_\lin{tot}$ is the sum of intrinsic processes $\alpha_\lin{int}$, spin pumping into the Pt and Ta layers $\alpha_\lin{Pt(Ta)}$, and possible spin memory loss $\alpha_\lin{SML}$. 

\begin{equation}
\alpha_\lin{tot} = \alpha_\lin{int} + \alpha_\lin{Pt} + \alpha_\lin{Ta} + \alpha_\lin{SML}
\label{eq:totalDamping}
\end{equation}

We can apply Eq. \ref{eq:totalDamping} to each of the stacking orders (Py/Pt and Pt/Py), and use the damping measurements for Py/Cu and Cu/Py control samples as a measure of $\alpha_\lin{int} + \alpha_\lin{Ta}$ for Py/Pt and Pt/Py, respectively. We note that that the total Gilbert damping for the two stacking orders differs by only 8\% (see Table \ref{tab:FMRParams}), while the odd SOT conductivity differs by a factor of 4. This suggests that the damping-like procceses contributing to $\sigmaO$ (i.e. iSHE) add only a small amount of enhanced damping, while the majority of spin current pumped out of the FM is lost and not available for iSHE conversion \cite{rojas-sanchez_spin_2014}. If we therefore assume that $\alpha_\lin{SML}$ is identical for the two stacking orders, and that the difference in $\sigmaO$ for the two stacks is due entirely to a difference in spin-mixing conductance, such that $\alpha_\lin{Pt}$(Py/Pt) = $4 \alpha_\lin{Pt}$(Pt/Py), then the resulting system of equations is solvable for $\alpha_\lin{Pt}$(Py/Pt) and $\alpha_\lin{Pt}$(Pt/Py), as well as $\alpha_\lin{SML}$. Using the results, we can estimate the spin pumping efficiency factor $\epsilon \equiv \alpha_\lin{sp}/(\alpha_\lin{sp} + \alpha_\lin{SML})$. We find that only 33\% or 13\% of the spin current pumped through the Pt interface is available for iSHE conversion, for Py/Pt and Pt/Py samples respectively. 

A more rigorous calculation can be done to estimate $G_{\uparrow\downarrow}$, $\epsilon$, and $\theta_\lin{SH}$ by simultaneously fitting Eq. \ref{eq:StilesHaney} and Eq. \ref{eq:totalDamping} for the two stacking orders (using the corrected values $(\sigmaO)_\lin{corr}$ from Table \ref{tab:SpinChargeParams} and total damping values from Table \ref{tab:FMRParams}). To perform this optimization, we use the functional form for the spin pumping damping contributions as presented in Ref. \citenum{boone_spin_2013}, such that $\alpha_\lin{Pt(Ta)}$ depends on $\lambda_\lin{s}$, $G_{\uparrow\downarrow}$, and $\sigma$ in order to implement the spin current backflow correction. We obtained a value for the Pt charge conductivity $\sigma = $ \SI{4.16e6}{\ohm^{-1}\meter^{-1}} from four-probe resistance measurement on a series of Py/Pt samples with varying Pt thickness, to allow isolation of the Pt contribution to the total conductivity. Using a value of $\lambda_\lin{s}$ = \SI{3.4}{\nano\meter} from Ref. \citenum{rojas-sanchez_spin_2014}, we obtain a spin Hall angle of $\theta_\lin{SH}$ = 0.28. This falls within the range of published values from DC spin Hall measurements (0.01--0.33) \cite{isasa_temperature_2015, mosendz_detection_2010, morota_indication_2011, liu_spin-torque_2011, weiler_detection_2014, weiler_experimental_2013, weiler_phase-sensitive_2014, obstbaum_inverse_2014, pai_dependence_2015}. In good agreement with the estimate above, we find efficiencies of 34\% and 18\% for Py/Pt and Pt/Py respectively. Furthermore, this optimization yields $G_{\uparrow\downarrow} = $ \SI{8.9e14}{\ohm^{-1}\m^{-2}} (for Py/Pt) and \SI{2.3e14}{\ohm^{-1}\m^{-2}} (for Pt/Py). Both of these values are below the Sharvin conductance \cite{liu_interface_2014} ($G_{\uparrow\downarrow}=$\SI{1e15}{\ohm^{-1}\meter^{-2}}), which serves as the theoretical upper bound for the spin-mixing conductance. This result demonstrates clearly that when Py is deposited on Pt, the FM/NM interface is detrimental to spin transport.

\section{Conclusion}
In summary, we have quantified both field- and damping-like inverse spin-orbit torques in \Py/Pt bilayers using phase-sensitive VNA-FMR measurements and an analysis of the sample's complex inductance that arises in part from the AC currents due to spin-charge conversion. The magnitude of these currents is determined by their respective SOT conductivities, a key figure of merit for characterizating and optimizing operational spintronic devices. Because our technique entails straightforward post-measurement data processing for an experimental technique that is well-established in the field, it provides a powerful way to unpick a highly complex experimental system and represents a broadly applicable tool for studying strong SOC material systems. The technique could even be applied to previously-acquired VNA-FMR data sets in which only spectroscopic analysis was performed. The measurements presented here demonstrate that both Rashba-Edelstein and spin Hall effects must be considered in FM/NM metallic bilayers. Together with the observation of significant variation in $\sigmaO$ with respect to FM/NM stacking order, these results point to interfacial engineering as an opportunity for enhancing current-controlled magnetism.

\begin{acknowledgments}
The authors would like to thank Mark Stiles and Mark Keller for many helpful discussions and illuminating insights.
\end{acknowledgments}



\bibliographystyle{apsrev4-1}
\bibliography{Physics}

\clearpage

%% file: Inductive_detection_of_AC_charge_currents_v17.4_SI.tex
\begin{widetext}
\begin{center}
\textbf{\large Supplementary Information} 
\end{center}
\end{widetext}

\setcounter{section}{0}
\setcounter{equation}{0}
\setcounter{figure}{0}
\setcounter{table}{0}
\setcounter{page}{1}
\makeatletter
\renewcommand{\theequation}{S\arabic{equation}}
\renewcommand{\thefigure}{S\arabic{figure}}
\renewcommand{\thetable}{S\arabic{table}}
\renewcommand{\bibnumfmt}[1]{[S#1]}
\renewcommand{\citenumfont}[1]{S#1}

\section{Sample Fabrication}
\label{sec:sampleFab}

\begin{table}[h]
\centering
\begin{tabular}{l | l}
\toprule
Sample & Deposition Order \\
\hline
Py/Pt & Substrate/Ta(1.5)/Py(3.5)/Pt(6)/Ta(3)  \\
Pt/Py & Substrate/Ta(1.5)/Pt(6)/Py(3.5)/Ta(3) \\
Py/Cu & Substrate/Ta(1.5)/Py(3.5)/Cu(3.3)/Ta(3) \\
Cu/Py & Substrate/Ta(1.5)/Cu(3.3)/Py(3.5)/Ta(3) \\
\botrule
\end{tabular}
\caption{Sample deposition orders and metallization thicknesses (in nanometers).}
\label{tab:DepOrder}
\end{table}

All samples were prepared by DC magnetron sputtering in an Ar base pressure of $\approx$\SI{0.07}{\pascal} ($\approx$\SI{0.5}{\milli\torr}) and a chamber base pressure of \SI{3e-6}{\pascal} (\SI{2e-8}{\torr}) on 3-inch wafers of thermally oxidized (100) Si (nominal resistivity = \SI{3}{\ohm\centi\meter}). The wafers were rotated at \SI{1}{\hertz} to \SI{2}{\hertz} during deposition to eliminate growth-induced anisotropy, and the sample holder was held at room temperature. All samples were grown on a \SI{1.5}{\nano\meter} Ta seed layer to promote (111) textured growth, which was then followed by the FM/NM (or NM/FM) bilayer. X-ray diffraction shows that the Ta seed layer is unordered. A \SI{3}{\nano\meter} Ta cap layer prevents oxidation of the FM and NM layers. It is expected that \SI{1}{\nano\meter} to \SI{2}{\nano\meter} of the cap layer forms the insulator TaO when exposed to air. Deposition order and film thicknesses are shown in Table \ref{tab:DepOrder}. The Pt and Cu thicknesses were chosen so that the DC conductivities (as characterized by a four-probe measurement) of the sample and control were equal, to ensure equality of Faraday induced currents. The wafers were subsequently coated with \SI{8}{\micro\meter} of photoresist to provide electrical insulation from the CPW and reduce the capacitive coupling of the CPW to the metallic layers. The wafers were diced to precise sizes using an automatic dicing saw. 

\section{Magnetic Characterization}
\subsection{Magnetic Susceptibility}
For our geometry, the driving microwave magnetic field lies primarily along $\hat{y}$, and we are concerned with the AC component of magnetization along $\hat{y}$ (see Fig. \ref{fig:CPWandCoords} in the main text for coordinate system). Therefore, the $S_{21}$ spectra are fit to the ${\chi}_{yy}$ component of the complex magnetic Polder susceptibility tensor in order to extract resonance field, linewidth, amplitude, and phase. 


\begin{equation}
\begin{bmatrix}
M_x \\
M_y
\end{bmatrix} =
\begin{bmatrix}
{\chi}_{xx} && {\chi}_{xy} \\
{\chi}_{yx} && {\chi}_{yy}
\end{bmatrix}
\begin{bmatrix}
h_x \\
h_y
\end{bmatrix}
\label{eq:linearSusc}
\end{equation}

\begin{widetext}
\begin{equation}
{\chi}(\omega, H_0) = \frac{M_\lin{s}}{\left( (H_0 - M_\lin{eff})^2 - \left(\dfrac{\omega}{\gamma \mu_0}\right)^2 + i \dfrac{2 \alpha_\lin{eff} \omega (H_0 - M_\lin{eff})}{\gamma \mu_0}\right)}
\begin{bmatrix}
(H_0 - M_\lin{eff}) && i\dfrac{\omega}{\gamma\mu_0} \\
-i\dfrac{\omega}{\gamma\mu_0} &&  (H_0 - M_\lin{eff})
\label{eq:chi}
\end{bmatrix}
\end{equation}
\end{widetext}

\noindent where $H_0$ is the externally applied DC field, $M_\lin{eff} = M_\lin{s} - H_\lin{k}^{\perp}$ is the effective magnetization, $M_\lin{s}$ is the saturation magnetization, $H_\lin{k}^{\perp}$ is the perpendicular anisotropy field, $\omega$ is the driving frequency, $\gamma$ is the gyromagnetic ratio, $\mu_0$ the vacuum permeability, and $\alpha_\lin{eff} = \alpha + \gamma \mu_0 \Delta H_0/(2\omega)$ is the effective damping parameter, with Gilbert damping constant $\alpha$ and inhomogeneous broadening $\Delta H_0$. 

The frequency dependence of the resonant field $H_\lin{res}$ and linewidth $\Delta H$ allow extraction of the effective magnetization $M_\lin{eff} = M_\lin{s} - H_\lin{k}^{\perp}$, spectroscopic g-factor $g$, inhomogeneous broadening $\Delta H_0$, and Gilbert damping parameter $\alpha$. We used SQUID magnetometry to measure the magnetization per unit area for all samples. Magnetization, g-factor, and damping values are summarized in Table \ref{tab:FMRParams}.

\subsection{Resonance Field Dispersion}
From the susceptibility fits to the $S_{21}$ spectra, we extract the resonance field as a function of microwave frequency. This is expected to follow the Kittel dispersion \cite{kittel_introduction_2004} for out-of-plane field $H_0$.

\begin{equation}
\omega = \mu_0 \gamma (H_\lin{res} - M_\lin{eff})
\end{equation}

\begin{figure}
	\centering
	\includegraphics[width=\linewidth]{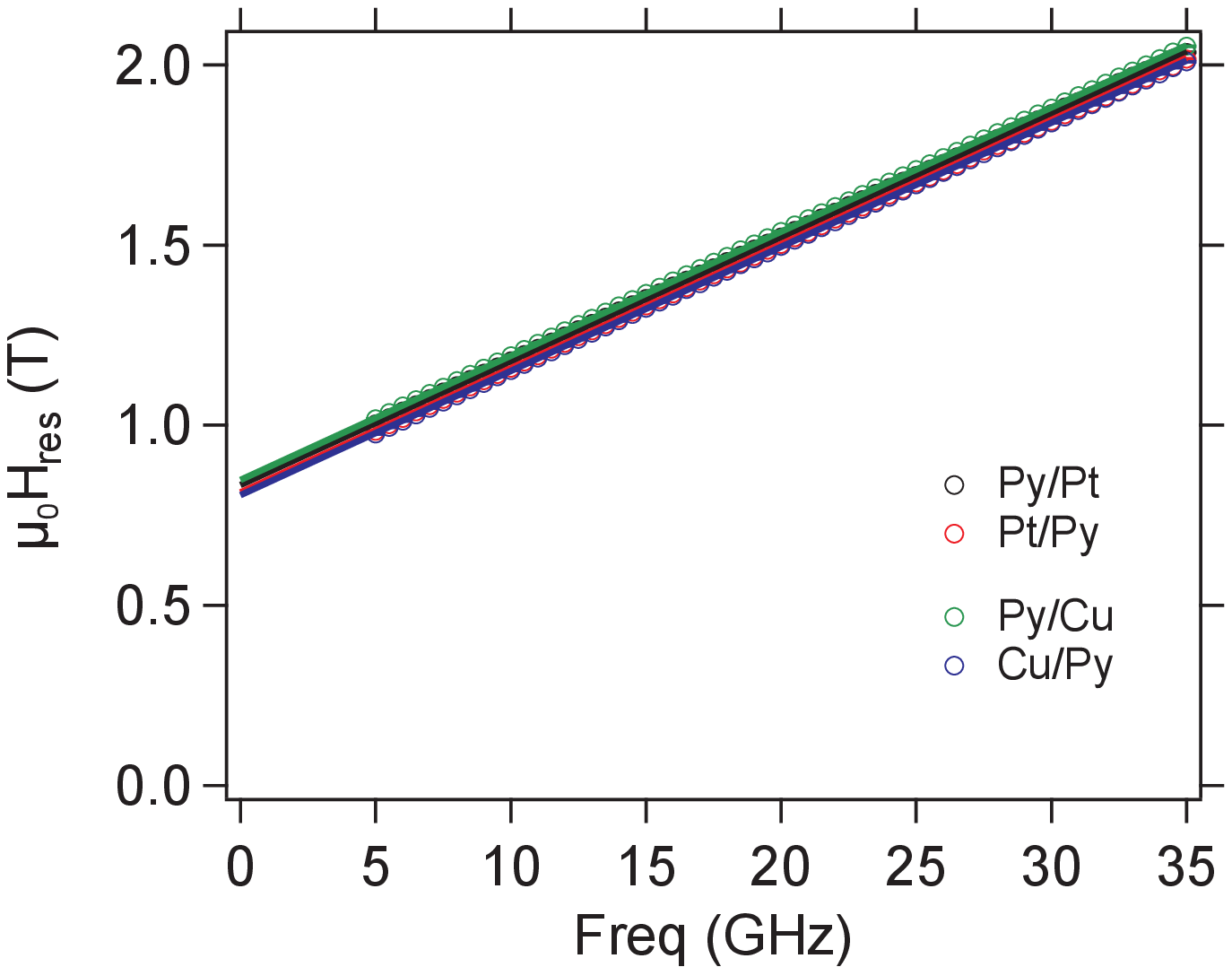}
	\caption{Resonance field vs. frequency dispersion, to extract spectroscopic g-factor, and $M_\lin{eff}$.}
	\label{fig:Hres_v_f}
\end{figure}

\noindent A plot of $\mu_0 H_\lin{res}$ vs. $f = \omega/2\pi$ is shown in Fig. \ref{fig:Hres_v_f}, with slope set by the gyromagnetic ratio $\gamma = g \mu_\lin{B}/\hbar$, and $y$-intercept set by $\mu_0 M_\lin{eff}$.

\subsection{Linewidth and Damping}
\label{sec:Damping}
The resonance linewidth is determined by the Gilbert damping constant $\alpha$ and inhomogeneous broadening $\Delta H_0$ according to

\begin{equation}
\mu_0 \Delta H = \mu_0 \Delta H_0 + \frac{2 \omega \alpha}{\gamma}
\label{eq:LW_v_f}
\end{equation}

\begin{figure}
	\centering
	\includegraphics[width=\linewidth]{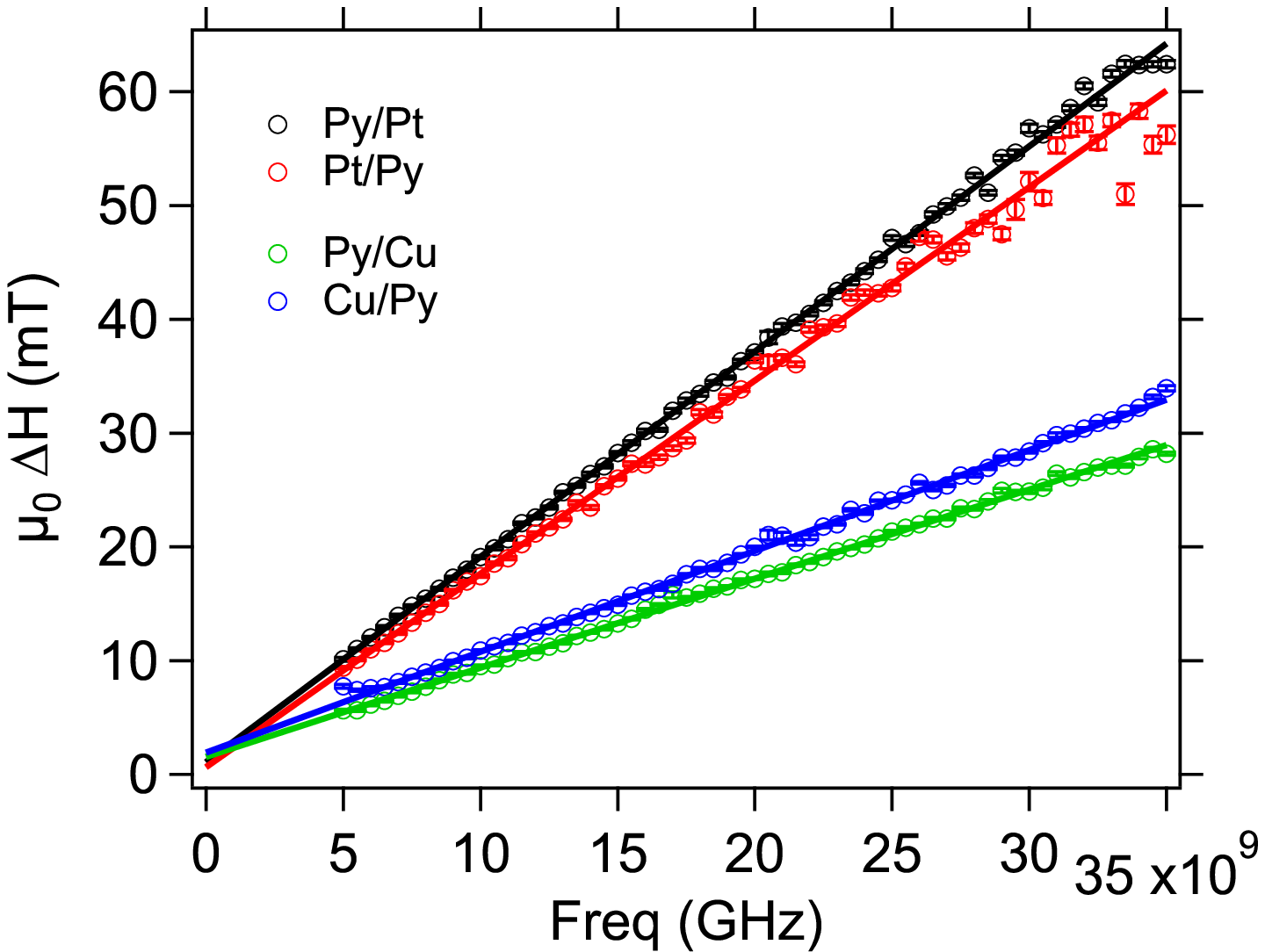}
	\caption{Resonance linewidth vs. frequency, to extract Gilbert damping constant $\alpha$ and inhomogeneous broadening.}
	\label{fig:dH_v_f}
\end{figure}

\noindent Data and fits of Eq. \ref{eq:LW_v_f} for the \SI{6}{\milli\meter} long samples for each deposition order are shown in Fig. \ref{fig:dH_v_f}. 

\begingroup
\squeezetable
\begin{table*}
\centering
\begin{tabular}{l*{6}{S}}
\toprule
Sample & {$M_\lin{eff}$ (\SI{}{\kilo\ampere/\meter})} & {g} & {$\mu_0 \Delta H_0 (\SI{}{\milli\tesla})$} & {$\alpha$} & {$M_\lin{s} d_\lin{FM}$ (\SI{}{\micro\ampere})} \\
\hline


Py/Pt & 663.5(7) & 2.079(1) & 1.2(8) & 0.0261(3) & 2069(1)  \\
Pt/Py & 647(1) & 2.079(3) & 2(2) & 0.0241(8) & 2121(1)  \\
Py/Cu & 674(1) & 2.075(1) & 1.1(5) & 0.0115(1) & 2341(2)  \\
Cu/Py & 642(1) & 2.077(1) & 1.7(9) & 0.0129(2) & 2077.0(4)   \\

\botrule
\end{tabular}
\caption{FMR and SQUID parameters for Py/Pt and Py/Cu bilayers.}
\label{tab:FMRParams}
\end{table*}
\endgroup

\subsection{SQUID Measurement}
We measured in-plane hysteresis curves at room temperature to determine the saturation moment of our samples. This total moment was normalized by the sample area to obtain $M_\lin{s} d_\lin{FM}$ (see Table \ref{tab:FMRParams}).

\section{Determination of Signal Phase}
\label{sec:SigPhase}
We consider the sample and CPW in a lumped element circuit model, in which the sample contributes an impedance $i \omega L$ to the circuit, in series with the characteristic impedance $Z_0$ of the CPW. Therefore, at the sample (or device-under-test), the current is simply given by:

\begin{eqnarray}
I_\lin{DUT} &=& \frac{V_1}{Z_0 + i \omega L} \nonumber \\
&\approx& \frac{V_1}{Z_0} \left(1 - \frac{i \omega L}{Z_0}\right) \\
&=& I_\lin{CPW} + \Delta I \nonumber
\end{eqnarray}

\noindent for $\omega L << Z_0$, and where $I_\lin{CPW}$ is the current in the unloaded CPW (with a positive Real current flowing in the $+\hat{x}$ direction). Therefore:

\begin{equation}
\Delta I = -\left(\frac{i \omega L}{Z_0}\right) I_\lin{CPW}
\label{eq:DeltaI}
\end{equation}

\noindent Using the dipolar inductance of Eq. \ref{eq:L0}, and considering the current response at the FMR condition, such that $\chi_{yy} = -i \gamma \mu_0 M_s /(2 \alpha_\lin{eff} \omega_\lin{res})$ (for CCW precession), we find:

\begin{equation}
\Delta I_\lin{dip} = -\frac{\gamma \mu_0^2 l \Msd \eta(z,W_\lin{wg})}{8 Z_0 \alpha_\lin{eff} W_\lin{wg}} I_\lin{CPW}
\label{eq:DeltaI_DUT}
\end{equation}

\begin{figure}
	\centering
	\includegraphics[width=\linewidth]{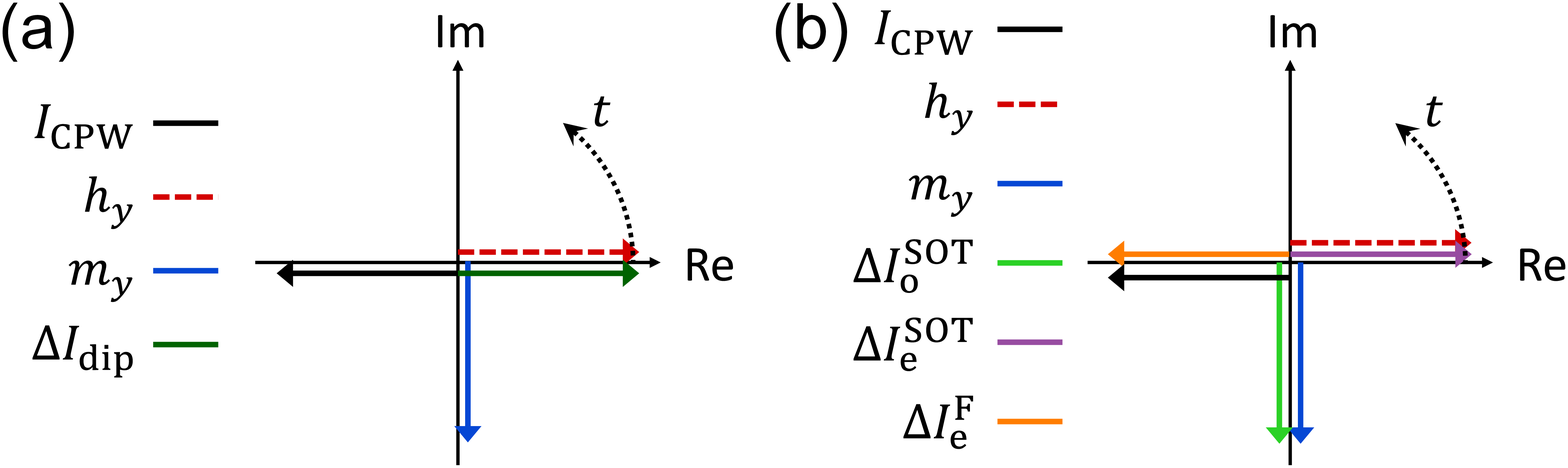}
	\caption{(a). Phasor diagram describing phase of current due to dipolar coupling to precessing magnetization $\mathbf{m}$, relative to $h_y$ at the FMR condition. The current $\Delta I_\lin{dip}$ creates a dissipative response. (b) Same as (a), but for currents in the CPW due to currents $I_\lin{NM}$ caused by Faraday and iSOT effects. Even currents appear dissipative or resistive, odd currents appear reactive. Note that all currents are defined such that a positive Real current in the CPW flows in the $+\hat{x}$ direction, and relative magnitudes are not indicated.}
\label{fig:CPWCurrentPhasor}
\end{figure}

From Eq. \ref{eq:DeltaI_DUT} we see that the change in current is in-phase with, but opposite in sign to the current responsible for $h_y$ (as depicted in Fig. \ref{fig:CPWCurrentPhasor}(a)). This change in current could be viewed as a change in the CPW resistance. That is, the sample inductance creates a purely dissipative response at the FMR condition, which is clearly seen in Fig. \ref{fig:S21_evolution}(a) and (b), and is expected for a spin system on resonance.

Let us now consider the phase of the currents in the CPW due to currents in the NM (from the Faraday and iSOT processes). These effects are captured by Fig. \ref{fig:CPWandCoords}(b-d) and the derivation of Sec. \ref{sec:L_NM}. For simplicity, we first focus on the Faraday-type currents in the NM. At time $t_0$, this current is maximum along the $\hat{x}$ direction. Via the mutual inductance between sample and CPW, an ``image current" flows in the CPW opposite to the Faraday current in the NM. Extending this logic to all current sources in the NM layer, we produce the phasor diagram of Fig. \ref{fig:CPWCurrentPhasor}(b). This demonstrates clearly that at the FMR condition, currents with even time-reversal symmetry create a dissipative response in the CPW, while odd-symmetry currents create a reactive response. The contribution of even and odd currents to dissipative or reactive response changes as field is swept through the resonance condition, resulting in the evolving lineshapes observed in Fig. \ref{fig:S21_evolution}(c) and (d). 

In order to coherently add the perturbative currents due to $L_0$ and $L_\lin{NM}$ to satisfy the above discussion (i.e. to combine the effects of Fig. \ref{fig:CPWCurrentPhasor}(a) and (b) with the proper phase assignment), we find:

\begin{eqnarray}
\Delta I_\lin{tot} &=& \Delta I_{L_0} + \Delta I_{L_\lin{NM}} \nonumber \\
&=& \left( -\frac{i \omega L_0}{Z_0} - \frac{\omega L_\lin{NM}}{Z_0}\right)I_\lin{CPW} \\
&=& -\left(\frac{i \omega L_\lin{tot}}{Z_0}\right) I_\lin{CPW}
\end{eqnarray}

\noindent where $L_\lin{tot} \equiv L_0 - i L_\lin{NM}$. Using this result, we recover the complex inductance relationships given by Eqs. \ref{eq:ReL} and \ref{eq:ImL}.

\section{Phase Error of $\Delta S_{21}$}
\label{sec:AnPhase}

\begin{figure}
	\centering
	\includegraphics[width=\linewidth]{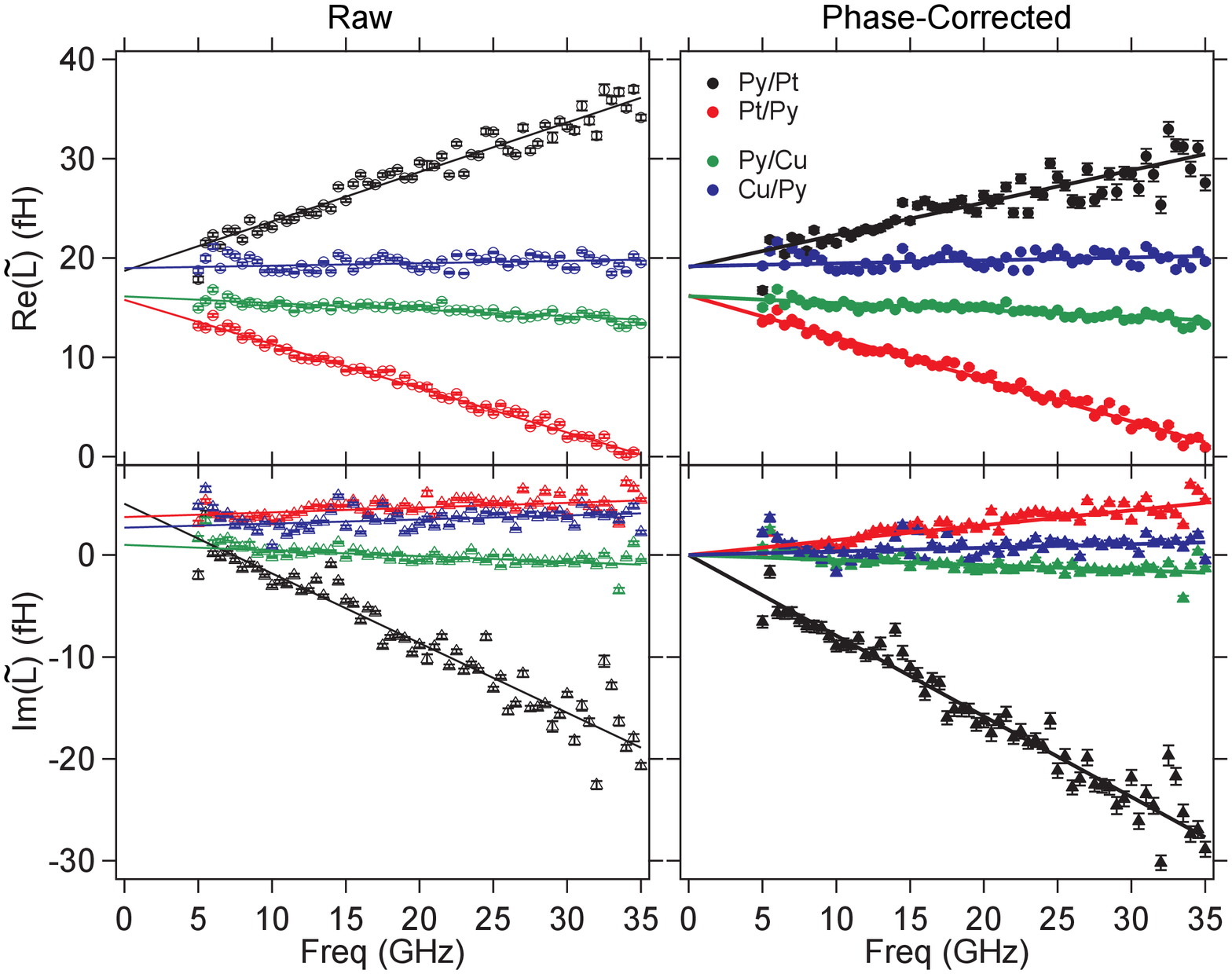}
	\caption{Correction of phase to enforce $\mbox{Im}(\tilde{L})(f=0) = 0$ for $l=$ \SI{6}{\milli\meter} sample. Raw data (left panels) show a small, non-zero component of $\mbox{Im}(\tilde{L})$ at $f=0$, which is unphysical. We therefore apply a small correction to eliminate this non-zero $y$-intercept, resulting in the phase-corrected data (right panels). }
	\label{fig:PhiAnom_Corr}
\end{figure}

\begin{table}
\begin{tabular}{c S}
\toprule
Sample & {$\phi_\lin{corr}$ (deg)} \\
\hline

Py/Pt & 12(1) \\
Pt/Py & 11.6(4)\\
Py/Cu & 1.8(8) \\
Cu/Py &  7.2(3) \\
\botrule
\end{tabular}
\caption{}
\label{tab:PhiCorr}
\end{table}

The background correction procedure of Sec. \ref{sec:BGCorr} requires one further phase correction in order to enforce that $\mbox{Im}(\tilde{L})(f=0) = 0$, as any finite $\mbox{Im}(\tilde{L})$ at zero frequency would be unphysical. However, as can be seen in the raw data of Fig. \ref{fig:PhiAnom_Corr}, the intercept of $\mbox{Im}(\tilde{L})$ at $f=0$ is indeed a small, finite number (left panels). In addition to the background phase correction described in Eq. \ref{eq:BGcorr_calc}, we therefore force an additional phase correction $\phi_\lin{corr} = \arctan[\mbox{Im}(\tilde{L})(f=0)/\mbox{Re}(\tilde{L})(f=0)]$. The $\phi_\lin{corr}$ necessary for each sample is shown in Table \ref{tab:PhiCorr}. 


\section{Shunting Correction}
\label{sec:ShuntCorr}
Our samples exhibit a shunting effect when the metallic thicknesses are such that the sheet resistance of the sample drops below \SI{50}{\ohm} ($Z_0$, the characteristic impedance of our CPW). This is similar to the shunting effect described in Ref. \citenum{jiao_spin_2013}. However, in that case, the attenuation of voltage signals as sample thickness increases follows immediately from Ohm's law and the decreasing resistance across which the iSHE voltage is measured. In our inductive measurements, the AC currents driven by iSOT generate signal voltages across the characteristic impedance of the CPW, $Z_0$. However, when the sample is thick enough, there is also a current return path through the thickness of the sample. For very thick samples, the integrated current through the sample thickness is zero (equal forward and return currents), and the inductive signal drops to zero. 

We therefore model the iSOT effects as a current source which drives current through parallel resistances $Z_0$ and $R_\lin{s}$, where $R_\lin{s}$ is the measured sheet resistance of our sample. For all samples in this study $R_\lin{s}$ was found to be $\approx$\SI{34}{\ohm}. In this model, only the fraction of the total current generated by iSOT that flows through the $Z_0$ branch can generate an inductive signal, corresponding to a fraction $R_\lin{s}/(Z_0 + R_\lin{s}) \approx 0.4$ of the total current. We therefore scale $\sigmaE$ and $\sigmaO$ by $\approx2.5$. Note that the Faraday effect acts as a source of $emf$, such that the currents due to the Faraday effect are observed to increase linearly with sample thickness, in accordance with Ohm's Law. Therefore, we do not correct $\sigmaF$ by the same shunting factor.

\section{Measurement of Permalloy resistivity}
\label{sec:IntResistivity}

\begin{figure}
	\centering
	\includegraphics[width=\linewidth]{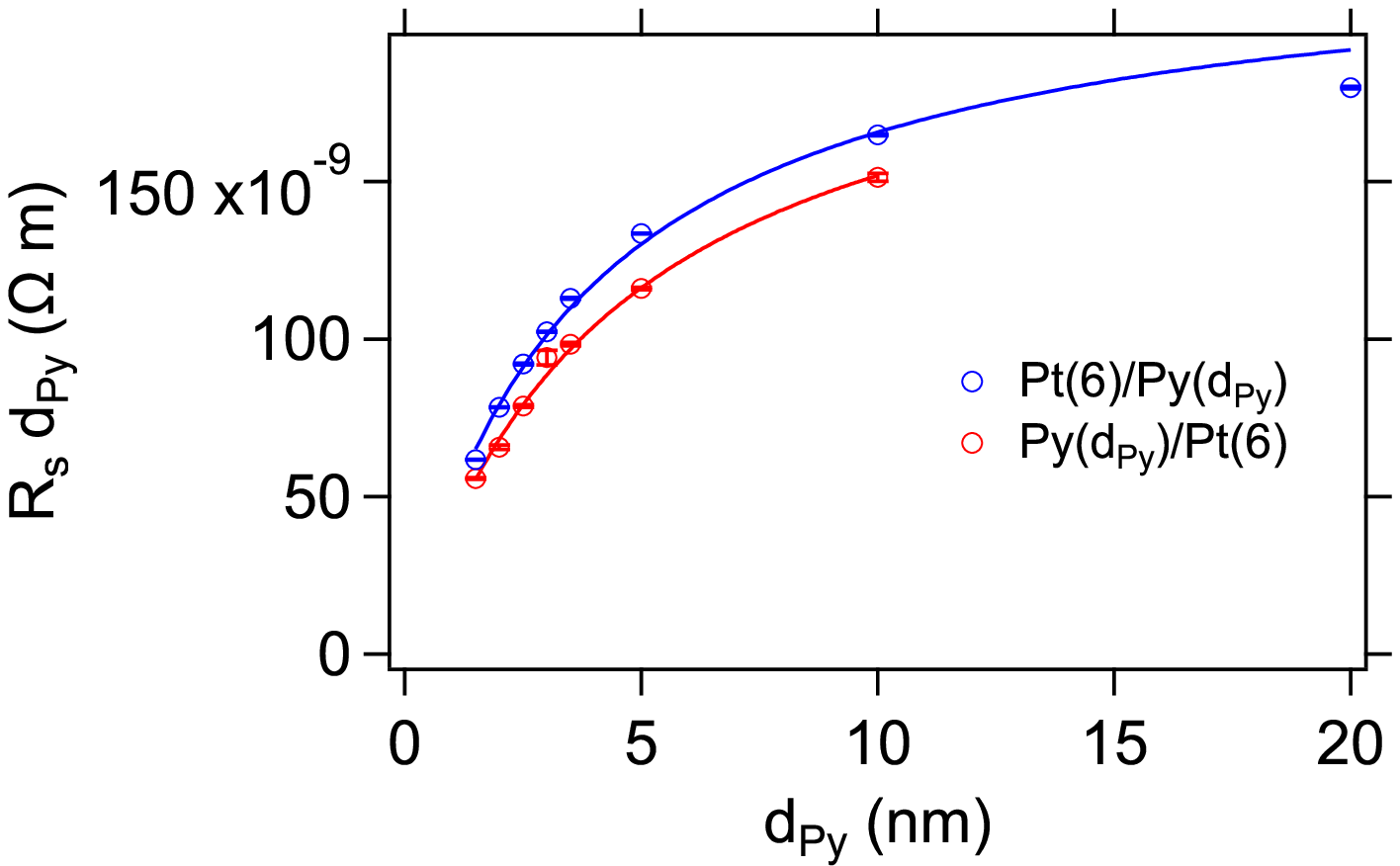}
	\caption{Measured sheet resistance vs. Py thickness $d_\lin{Py}$ for both stacking orders of Py and Pt: Ta(1.5)/Py($d_\lin{Py}$)/Pt(6)/Ta(1.5) and Ta(1.5)/Pt(6)/Py($d_\lin{Py}$)/Ta(1.5). Eq. \ref{eq:Rd_fit} is used as the fit function.}
	\label{fig:Rd_Py}
\end{figure}

In order to determine the interface conductivity $\sigma_\lin{int}$ used for determination of $\alpha_\lin{R}$ in main text Eq. \ref{eq:alpha_Rashba}, we measured the resistivity of Ta(1.5)/Py($d_\lin{Py}$)/Pt(6)/Ta(3) and Ta(1.5)/Pt(6)/Py($d_\lin{Py}$)/Ta(3) films (thicknesses in nanometers) as a function of Py film thickness, $d_\lin{Py}$ (Fig. \ref{fig:Rd_Py}). In each case, we find that the data are well-described by a simple model in which the Py resistivity is independent of thickness, and adds as a parallel resistance with the Pt and Ta conducting layers. That is, the total sheet resistance $R_\lin{s}$ is given by: $1/R_\lin{s} = d_\lin{Py}/\rho_0 + 1/R_\lin{other}$, where $\rho_0$ is the Py bulk resistivity, and $R_\lin{other}$ is the combined sheet resistance of the Pt and Ta layers. We multiply the measured sheet resistance by the Py thickness, such that

\begin{equation}
R_\lin{s} d_\lin{Py} = \frac{d_\lin{Py}}{\dfrac{d_\lin{Py}}{\rho_0} + \dfrac{1}{R_\lin{other}}}
\label{eq:Rd_fit}
\end{equation}

From the fits shown in Fig. \ref{fig:Rd_Py}, we find $\rho_0 = $ \SI{21.9(2)e-8}{\ohm\meter} and $R_\lin{other} = $ \SI{49.5(4)}{\ohm} for the Py/Pt sample, and , $\rho_0 = $ \SI{22.78(4)e-8}{\ohm\meter} and $R_\lin{other} = $ \SI{60.7(1)}{\ohm} for the Pt/Py sample. To calulate $\sigma_\lin{int}$ for  Eq. \ref{eq:alpha_Rashba}, we simply use the inverse of these bulk resistivity values.

\bibliography{Physics}
\bibliographystyle{apsrev4-1}

\clearpage

%% file: main_17.4.bbl
\begin{thebibliography}{51}%
\makeatletter
\providecommand \@ifxundefined [1]{%
 \@ifx{#1\undefined}
}%
\providecommand \@ifnum [1]{%
 \ifnum #1\expandafter \@firstoftwo
 \else \expandafter \@secondoftwo
 \fi
}%
\providecommand \@ifx [1]{%
 \ifx #1\expandafter \@firstoftwo
 \else \expandafter \@secondoftwo
 \fi
}%
\providecommand \natexlab [1]{#1}%
\providecommand \enquote  [1]{``#1''}%
\providecommand \bibnamefont  [1]{#1}%
\providecommand \bibfnamefont [1]{#1}%
\providecommand \citenamefont [1]{#1}%
\providecommand \href@noop [0]{\@secondoftwo}%
\providecommand \href [0]{\begingroup \@sanitize@url \@href}%
\providecommand \@href[1]{\@@startlink{#1}\@@href}%
\providecommand \@@href[1]{\endgroup#1\@@endlink}%
\providecommand \@sanitize@url [0]{\catcode `\\12\catcode `\$12\catcode
  `\&12\catcode `\#12\catcode `\^12\catcode `\_12\catcode `\%12\relax}%
\providecommand \@@startlink[1]{}%
\providecommand \@@endlink[0]{}%
\providecommand \url  [0]{\begingroup\@sanitize@url \@url }%
\providecommand \@url [1]{\endgroup\@href {#1}{\urlprefix }}%
\providecommand \urlprefix  [0]{URL }%
\providecommand \Eprint [0]{\href }%
\providecommand \doibase [0]{http://dx.doi.org/}%
\providecommand \selectlanguage [0]{\@gobble}%
\providecommand \bibinfo  [0]{\@secondoftwo}%
\providecommand \bibfield  [0]{\@secondoftwo}%
\providecommand \translation [1]{[#1]}%
\providecommand \BibitemOpen [0]{}%
\providecommand \bibitemStop [0]{}%
\providecommand \bibitemNoStop [0]{.\EOS\space}%
\providecommand \EOS [0]{\spacefactor3000\relax}%
\providecommand \BibitemShut  [1]{\csname bibitem#1\endcsname}%
\let\auto@bib@innerbib\@empty
\bibitem [{\citenamefont {Freimuth}\ \emph {et~al.}(2015)\citenamefont
  {Freimuth}, \citenamefont {Blügel},\ and\ \citenamefont
  {Mokrousov}}]{freimuth_direct_2015}%
  \BibitemOpen
  \bibfield  {author} {\bibinfo {author} {\bibfnamefont {F.}~\bibnamefont
  {Freimuth}}, \bibinfo {author} {\bibfnamefont {S.}~\bibnamefont {Blügel}}, \
  and\ \bibinfo {author} {\bibfnamefont {Y.}~\bibnamefont {Mokrousov}},\ }\href
  {\doibase 10.1103/PhysRevB.92.064415} {\bibfield  {journal} {\bibinfo
  {journal} {Physical Review B}\ }\textbf {\bibinfo {volume} {92}},\ \bibinfo
  {pages} {064415} (\bibinfo {year} {2015})}\BibitemShut {NoStop}%
\bibitem [{\citenamefont {Mihai~Miron}\ \emph {et~al.}(2010)\citenamefont
  {Mihai~Miron}, \citenamefont {Gaudin}, \citenamefont {Auffret}, \citenamefont
  {Rodmacq}, \citenamefont {Schuhl}, \citenamefont {Pizzini}, \citenamefont
  {Vogel},\ and\ \citenamefont
  {Gambardella}}]{mihai_miron_current-driven_2010}%
  \BibitemOpen
  \bibfield  {author} {\bibinfo {author} {\bibfnamefont {I.}~\bibnamefont
  {Mihai~Miron}}, \bibinfo {author} {\bibfnamefont {G.}~\bibnamefont {Gaudin}},
  \bibinfo {author} {\bibfnamefont {S.}~\bibnamefont {Auffret}}, \bibinfo
  {author} {\bibfnamefont {B.}~\bibnamefont {Rodmacq}}, \bibinfo {author}
  {\bibfnamefont {A.}~\bibnamefont {Schuhl}}, \bibinfo {author} {\bibfnamefont
  {S.}~\bibnamefont {Pizzini}}, \bibinfo {author} {\bibfnamefont
  {J.}~\bibnamefont {Vogel}}, \ and\ \bibinfo {author} {\bibfnamefont
  {P.}~\bibnamefont {Gambardella}},\ }\href {\doibase 10.1038/nmat2613}
  {\bibfield  {journal} {\bibinfo  {journal} {Nature Materials}\ }\textbf
  {\bibinfo {volume} {9}},\ \bibinfo {pages} {230} (\bibinfo {year}
  {2010})}\BibitemShut {NoStop}%
\bibitem [{\citenamefont {Duan}\ \emph {et~al.}(2014)\citenamefont {Duan},
  \citenamefont {Smith}, \citenamefont {Yang}, \citenamefont {Youngblood},
  \citenamefont {Lindner}, \citenamefont {Demidov}, \citenamefont
  {Demokritov},\ and\ \citenamefont {Krivorotov}}]{duan_nanowire_2014}%
  \BibitemOpen
  \bibfield  {author} {\bibinfo {author} {\bibfnamefont {Z.}~\bibnamefont
  {Duan}}, \bibinfo {author} {\bibfnamefont {A.}~\bibnamefont {Smith}},
  \bibinfo {author} {\bibfnamefont {L.}~\bibnamefont {Yang}}, \bibinfo {author}
  {\bibfnamefont {B.}~\bibnamefont {Youngblood}}, \bibinfo {author}
  {\bibfnamefont {J.}~\bibnamefont {Lindner}}, \bibinfo {author} {\bibfnamefont
  {V.~E.}\ \bibnamefont {Demidov}}, \bibinfo {author} {\bibfnamefont {S.~O.}\
  \bibnamefont {Demokritov}}, \ and\ \bibinfo {author} {\bibfnamefont {I.~N.}\
  \bibnamefont {Krivorotov}},\ }\href {\doibase 10.1038/ncomms6616} {\bibfield
  {journal} {\bibinfo  {journal} {Nature Communications}\ }\textbf {\bibinfo
  {volume} {5}} (\bibinfo {year} {2014}),\ 10.1038/ncomms6616}\BibitemShut
  {NoStop}%
\bibitem [{\citenamefont {Liu}\ \emph {et~al.}(2012)\citenamefont {Liu},
  \citenamefont {Pai}, \citenamefont {Li}, \citenamefont {Tseng}, \citenamefont
  {Ralph},\ and\ \citenamefont {Buhrman}}]{liu_spin-torque_2012}%
  \BibitemOpen
  \bibfield  {author} {\bibinfo {author} {\bibfnamefont {L.}~\bibnamefont
  {Liu}}, \bibinfo {author} {\bibfnamefont {C.-F.}\ \bibnamefont {Pai}},
  \bibinfo {author} {\bibfnamefont {Y.}~\bibnamefont {Li}}, \bibinfo {author}
  {\bibfnamefont {H.~W.}\ \bibnamefont {Tseng}}, \bibinfo {author}
  {\bibfnamefont {D.~C.}\ \bibnamefont {Ralph}}, \ and\ \bibinfo {author}
  {\bibfnamefont {R.~A.}\ \bibnamefont {Buhrman}},\ }\href {\doibase
  10.1126/science.1218197} {\bibfield  {journal} {\bibinfo  {journal}
  {Science}\ }\textbf {\bibinfo {volume} {336}},\ \bibinfo {pages} {555}
  (\bibinfo {year} {2012})}\BibitemShut {NoStop}%
\bibitem [{\citenamefont {Manchon}\ \emph {et~al.}(2015)\citenamefont
  {Manchon}, \citenamefont {Koo}, \citenamefont {Nitta}, \citenamefont
  {Frolov},\ and\ \citenamefont {Duine}}]{manchon_new_2015}%
  \BibitemOpen
  \bibfield  {author} {\bibinfo {author} {\bibfnamefont {A.}~\bibnamefont
  {Manchon}}, \bibinfo {author} {\bibfnamefont {H.~C.}\ \bibnamefont {Koo}},
  \bibinfo {author} {\bibfnamefont {J.}~\bibnamefont {Nitta}}, \bibinfo
  {author} {\bibfnamefont {S.~M.}\ \bibnamefont {Frolov}}, \ and\ \bibinfo
  {author} {\bibfnamefont {R.~A.}\ \bibnamefont {Duine}},\ }\href {\doibase
  10.1038/nmat4360} {\bibfield  {journal} {\bibinfo  {journal} {Nature
  Materials}\ }\textbf {\bibinfo {volume} {14}},\ \bibinfo {pages} {871}
  (\bibinfo {year} {2015})}\BibitemShut {NoStop}%
\bibitem [{\citenamefont {Czeschka}\ \emph {et~al.}(2011)\citenamefont
  {Czeschka}, \citenamefont {Dreher}, \citenamefont {Brandt}, \citenamefont
  {Weiler}, \citenamefont {Althammer}, \citenamefont {Imort}, \citenamefont
  {Reiss}, \citenamefont {Thomas}, \citenamefont {Schoch}, \citenamefont
  {Limmer}, \citenamefont {Huebl}, \citenamefont {Gross},\ and\ \citenamefont
  {Goennenwein}}]{czeschka_scaling_2011}%
  \BibitemOpen
  \bibfield  {author} {\bibinfo {author} {\bibfnamefont {F.~D.}\ \bibnamefont
  {Czeschka}}, \bibinfo {author} {\bibfnamefont {L.}~\bibnamefont {Dreher}},
  \bibinfo {author} {\bibfnamefont {M.~S.}\ \bibnamefont {Brandt}}, \bibinfo
  {author} {\bibfnamefont {M.}~\bibnamefont {Weiler}}, \bibinfo {author}
  {\bibfnamefont {M.}~\bibnamefont {Althammer}}, \bibinfo {author}
  {\bibfnamefont {I.-M.}\ \bibnamefont {Imort}}, \bibinfo {author}
  {\bibfnamefont {G.}~\bibnamefont {Reiss}}, \bibinfo {author} {\bibfnamefont
  {A.}~\bibnamefont {Thomas}}, \bibinfo {author} {\bibfnamefont
  {W.}~\bibnamefont {Schoch}}, \bibinfo {author} {\bibfnamefont
  {W.}~\bibnamefont {Limmer}}, \bibinfo {author} {\bibfnamefont
  {H.}~\bibnamefont {Huebl}}, \bibinfo {author} {\bibfnamefont
  {R.}~\bibnamefont {Gross}}, \ and\ \bibinfo {author} {\bibfnamefont
  {S.~T.~B.}\ \bibnamefont {Goennenwein}},\ }\href {\doibase
  10.1103/PhysRevLett.107.046601} {\bibfield  {journal} {\bibinfo  {journal}
  {Physical Review Letters}\ }\textbf {\bibinfo {volume} {107}},\ \bibinfo
  {pages} {046601} (\bibinfo {year} {2011})}\BibitemShut {NoStop}%
\bibitem [{\citenamefont {Weiler}\ \emph
  {et~al.}(2014{\natexlab{a}})\citenamefont {Weiler}, \citenamefont {Shaw},
  \citenamefont {Nembach},\ and\ \citenamefont
  {Silva}}]{weiler_detection_2014}%
  \BibitemOpen
  \bibfield  {author} {\bibinfo {author} {\bibfnamefont {M.}~\bibnamefont
  {Weiler}}, \bibinfo {author} {\bibfnamefont {J.~M.}\ \bibnamefont {Shaw}},
  \bibinfo {author} {\bibfnamefont {H.~T.}\ \bibnamefont {Nembach}}, \ and\
  \bibinfo {author} {\bibfnamefont {T.~J.}\ \bibnamefont {Silva}},\ }\href
  {\doibase 10.1109/LMAG.2014.2361791} {\bibfield  {journal} {\bibinfo
  {journal} {IEEE Magnetics Letters}\ }\textbf {\bibinfo {volume} {5}},\
  \bibinfo {pages} {3700104} (\bibinfo {year}
  {2014}{\natexlab{a}})}\BibitemShut {NoStop}%
\bibitem [{\citenamefont {Wang}\ \emph {et~al.}(2014)\citenamefont {Wang},
  \citenamefont {Du}, \citenamefont {Pu}, \citenamefont {Adur}, \citenamefont
  {Hammel},\ and\ \citenamefont {Yang}}]{wang_scaling_2014}%
  \BibitemOpen
  \bibfield  {author} {\bibinfo {author} {\bibfnamefont {H.}~\bibnamefont
  {Wang}}, \bibinfo {author} {\bibfnamefont {C.}~\bibnamefont {Du}}, \bibinfo
  {author} {\bibfnamefont {Y.}~\bibnamefont {Pu}}, \bibinfo {author}
  {\bibfnamefont {R.}~\bibnamefont {Adur}}, \bibinfo {author} {\bibfnamefont
  {P.}~\bibnamefont {Hammel}}, \ and\ \bibinfo {author} {\bibfnamefont
  {F.}~\bibnamefont {Yang}},\ }\href {\doibase 10.1103/PhysRevLett.112.197201}
  {\bibfield  {journal} {\bibinfo  {journal} {Physical Review Letters}\
  }\textbf {\bibinfo {volume} {112}},\ \bibinfo {pages} {197201} (\bibinfo
  {year} {2014})}\BibitemShut {NoStop}%
\bibitem [{\citenamefont {Garello}\ \emph {et~al.}(2013)\citenamefont
  {Garello}, \citenamefont {Miron}, \citenamefont {Avci}, \citenamefont
  {Freimuth}, \citenamefont {Mokrousov}, \citenamefont {Blügel}, \citenamefont
  {Auffret}, \citenamefont {Boulle}, \citenamefont {Gaudin},\ and\
  \citenamefont {Gambardella}}]{garello_symmetry_2013}%
  \BibitemOpen
  \bibfield  {author} {\bibinfo {author} {\bibfnamefont {K.}~\bibnamefont
  {Garello}}, \bibinfo {author} {\bibfnamefont {I.~M.}\ \bibnamefont {Miron}},
  \bibinfo {author} {\bibfnamefont {C.~O.}\ \bibnamefont {Avci}}, \bibinfo
  {author} {\bibfnamefont {F.}~\bibnamefont {Freimuth}}, \bibinfo {author}
  {\bibfnamefont {Y.}~\bibnamefont {Mokrousov}}, \bibinfo {author}
  {\bibfnamefont {S.}~\bibnamefont {Blügel}}, \bibinfo {author} {\bibfnamefont
  {S.}~\bibnamefont {Auffret}}, \bibinfo {author} {\bibfnamefont
  {O.}~\bibnamefont {Boulle}}, \bibinfo {author} {\bibfnamefont
  {G.}~\bibnamefont {Gaudin}}, \ and\ \bibinfo {author} {\bibfnamefont
  {P.}~\bibnamefont {Gambardella}},\ }\href {\doibase 10.1038/nnano.2013.145}
  {\bibfield  {journal} {\bibinfo  {journal} {Nature Nanotechnology}\ }\textbf
  {\bibinfo {volume} {8}},\ \bibinfo {pages} {587} (\bibinfo {year}
  {2013})}\BibitemShut {NoStop}%
\bibitem [{\citenamefont {Avci}\ \emph {et~al.}(2014)\citenamefont {Avci},
  \citenamefont {Garello}, \citenamefont {Nistor}, \citenamefont {Godey},
  \citenamefont {Ballesteros}, \citenamefont {Mugarza}, \citenamefont {Barla},
  \citenamefont {Valvidares}, \citenamefont {Pellegrin}, \citenamefont {Ghosh},
  \citenamefont {Miron}, \citenamefont {Boulle}, \citenamefont {Auffret},
  \citenamefont {Gaudin},\ and\ \citenamefont
  {Gambardella}}]{avci_fieldlike_2014}%
  \BibitemOpen
  \bibfield  {author} {\bibinfo {author} {\bibfnamefont {C.~O.}\ \bibnamefont
  {Avci}}, \bibinfo {author} {\bibfnamefont {K.}~\bibnamefont {Garello}},
  \bibinfo {author} {\bibfnamefont {C.}~\bibnamefont {Nistor}}, \bibinfo
  {author} {\bibfnamefont {S.}~\bibnamefont {Godey}}, \bibinfo {author}
  {\bibfnamefont {B.}~\bibnamefont {Ballesteros}}, \bibinfo {author}
  {\bibfnamefont {A.}~\bibnamefont {Mugarza}}, \bibinfo {author} {\bibfnamefont
  {A.}~\bibnamefont {Barla}}, \bibinfo {author} {\bibfnamefont
  {M.}~\bibnamefont {Valvidares}}, \bibinfo {author} {\bibfnamefont
  {E.}~\bibnamefont {Pellegrin}}, \bibinfo {author} {\bibfnamefont
  {A.}~\bibnamefont {Ghosh}}, \bibinfo {author} {\bibfnamefont {I.~M.}\
  \bibnamefont {Miron}}, \bibinfo {author} {\bibfnamefont {O.}~\bibnamefont
  {Boulle}}, \bibinfo {author} {\bibfnamefont {S.}~\bibnamefont {Auffret}},
  \bibinfo {author} {\bibfnamefont {G.}~\bibnamefont {Gaudin}}, \ and\ \bibinfo
  {author} {\bibfnamefont {P.}~\bibnamefont {Gambardella}},\ }\href {\doibase
  10.1103/PhysRevB.89.214419} {\bibfield  {journal} {\bibinfo  {journal}
  {Physical Review B}\ }\textbf {\bibinfo {volume} {89}},\ \bibinfo {pages}
  {214419} (\bibinfo {year} {2014})}\BibitemShut {NoStop}%
\bibitem [{\citenamefont {Fan}\ \emph {et~al.}(2014)\citenamefont {Fan},
  \citenamefont {Celik}, \citenamefont {Wu}, \citenamefont {Ni}, \citenamefont
  {Lee}, \citenamefont {Lorenz},\ and\ \citenamefont
  {Xiao}}]{fan_quantifying_2014}%
  \BibitemOpen
  \bibfield  {author} {\bibinfo {author} {\bibfnamefont {X.}~\bibnamefont
  {Fan}}, \bibinfo {author} {\bibfnamefont {H.}~\bibnamefont {Celik}}, \bibinfo
  {author} {\bibfnamefont {J.}~\bibnamefont {Wu}}, \bibinfo {author}
  {\bibfnamefont {C.}~\bibnamefont {Ni}}, \bibinfo {author} {\bibfnamefont
  {K.-J.}\ \bibnamefont {Lee}}, \bibinfo {author} {\bibfnamefont {V.~O.}\
  \bibnamefont {Lorenz}}, \ and\ \bibinfo {author} {\bibfnamefont {J.~Q.}\
  \bibnamefont {Xiao}},\ }\href {\doibase 10.1038/ncomms4042} {\bibfield
  {journal} {\bibinfo  {journal} {Nature Communications}\ }\textbf {\bibinfo
  {volume} {5}} (\bibinfo {year} {2014}),\ 10.1038/ncomms4042}\BibitemShut
  {NoStop}%
\bibitem [{\citenamefont {Weiler}\ \emph
  {et~al.}(2014{\natexlab{b}})\citenamefont {Weiler}, \citenamefont {Shaw},
  \citenamefont {Nembach},\ and\ \citenamefont
  {Silva}}]{weiler_phase-sensitive_2014}%
  \BibitemOpen
  \bibfield  {author} {\bibinfo {author} {\bibfnamefont {M.}~\bibnamefont
  {Weiler}}, \bibinfo {author} {\bibfnamefont {J.~M.}\ \bibnamefont {Shaw}},
  \bibinfo {author} {\bibfnamefont {H.~T.}\ \bibnamefont {Nembach}}, \ and\
  \bibinfo {author} {\bibfnamefont {T.~J.}\ \bibnamefont {Silva}},\ }\href
  {\doibase 10.1103/PhysRevLett.113.157204} {\bibfield  {journal} {\bibinfo
  {journal} {Physical Review Letters}\ }\textbf {\bibinfo {volume} {113}},\
  \bibinfo {pages} {157204} (\bibinfo {year} {2014}{\natexlab{b}})}\BibitemShut
  {NoStop}%
\bibitem [{\citenamefont {Onsager}(1931)}]{onsager_reciprocal_1931}%
  \BibitemOpen
  \bibfield  {author} {\bibinfo {author} {\bibfnamefont {L.}~\bibnamefont
  {Onsager}},\ }\href {\doibase 10.1103/PhysRev.37.405} {\bibfield  {journal}
  {\bibinfo  {journal} {Physical Review}\ }\textbf {\bibinfo {volume} {37}},\
  \bibinfo {pages} {405} (\bibinfo {year} {1931})}\BibitemShut {NoStop}%
\bibitem [{\citenamefont {Ramo}\ \emph {et~al.}(2008)\citenamefont {Ramo},
  \citenamefont {Whinnery},\ and\ \citenamefont {Duzer}}]{ramo_fields_2008}%
  \BibitemOpen
  \bibfield  {author} {\bibinfo {author} {\bibfnamefont {S.}~\bibnamefont
  {Ramo}}, \bibinfo {author} {\bibfnamefont {J.~R.}\ \bibnamefont {Whinnery}},
  \ and\ \bibinfo {author} {\bibfnamefont {T.~V.}\ \bibnamefont {Duzer}},\
  }\href@noop {} {{\selectlanguage {English}\emph {\bibinfo {title} {{FIELDS}
  {AND} {WAVES} {IN} {COMMUNICATION} {ELECTRONICS}, 3RD {ED}}}}}\ (\bibinfo
  {publisher} {Wiley-India},\ \bibinfo {year} {2008})\ \bibinfo {note}
  {google-Books-ID: VFFqU7pXl6gC}\BibitemShut {NoStop}%
\bibitem [{\citenamefont {White}(1985)}]{white_introduction_1985}%
  \BibitemOpen
  \bibfield  {author} {\bibinfo {author} {\bibfnamefont {R.~M.}\ \bibnamefont
  {White}},\ }\href@noop {} {{\selectlanguage {English}\emph {\bibinfo {title}
  {Introduction to magnetic recording}}}}\ (\bibinfo  {publisher} {IEEE
  Press},\ \bibinfo {year} {1985})\ \bibinfo {note} {google-Books-ID:
  4ydTAAAAMAAJ}\BibitemShut {NoStop}%
\bibitem [{\citenamefont {Saitoh}\ \emph {et~al.}(2006)\citenamefont {Saitoh},
  \citenamefont {Ueda}, \citenamefont {Miyajima},\ and\ \citenamefont
  {Tatara}}]{saitoh_conversion_2006}%
  \BibitemOpen
  \bibfield  {author} {\bibinfo {author} {\bibfnamefont {E.}~\bibnamefont
  {Saitoh}}, \bibinfo {author} {\bibfnamefont {M.}~\bibnamefont {Ueda}},
  \bibinfo {author} {\bibfnamefont {H.}~\bibnamefont {Miyajima}}, \ and\
  \bibinfo {author} {\bibfnamefont {G.}~\bibnamefont {Tatara}},\ }\href
  {\doibase doi:10.1063/1.2199473} {\bibfield  {journal} {\bibinfo  {journal}
  {Applied Physics Letters}\ }\textbf {\bibinfo {volume} {88}},\ \bibinfo
  {pages} {182509} (\bibinfo {year} {2006})}\BibitemShut {NoStop}%
\bibitem [{\citenamefont {Sánchez}\ \emph {et~al.}(2013)\citenamefont
  {Sánchez}, \citenamefont {Vila}, \citenamefont {Desfonds}, \citenamefont
  {Gambarelli}, \citenamefont {Attané}, \citenamefont {De~Teresa},
  \citenamefont {Magén},\ and\ \citenamefont
  {Fert}}]{sanchez_spin--charge_2013}%
  \BibitemOpen
  \bibfield  {author} {\bibinfo {author} {\bibfnamefont {J.~C.~R.}\
  \bibnamefont {Sánchez}}, \bibinfo {author} {\bibfnamefont {L.}~\bibnamefont
  {Vila}}, \bibinfo {author} {\bibfnamefont {G.}~\bibnamefont {Desfonds}},
  \bibinfo {author} {\bibfnamefont {S.}~\bibnamefont {Gambarelli}}, \bibinfo
  {author} {\bibfnamefont {J.~P.}\ \bibnamefont {Attané}}, \bibinfo {author}
  {\bibfnamefont {J.~M.}\ \bibnamefont {De~Teresa}}, \bibinfo {author}
  {\bibfnamefont {C.}~\bibnamefont {Magén}}, \ and\ \bibinfo {author}
  {\bibfnamefont {A.}~\bibnamefont {Fert}},\ }\href {\doibase
  10.1038/ncomms3944} {\bibfield  {journal} {\bibinfo  {journal} {Nature
  Communications}\ }\textbf {\bibinfo {volume} {4}},\ \bibinfo {pages} {2944}
  (\bibinfo {year} {2013})}\BibitemShut {NoStop}%
\bibitem [{\citenamefont {Katine}\ \emph {et~al.}(2000)\citenamefont {Katine},
  \citenamefont {Albert}, \citenamefont {Buhrman}, \citenamefont {Myers},\ and\
  \citenamefont {Ralph}}]{katine_current-driven_2000}%
  \BibitemOpen
  \bibfield  {author} {\bibinfo {author} {\bibfnamefont {J.~A.}\ \bibnamefont
  {Katine}}, \bibinfo {author} {\bibfnamefont {F.~J.}\ \bibnamefont {Albert}},
  \bibinfo {author} {\bibfnamefont {R.~A.}\ \bibnamefont {Buhrman}}, \bibinfo
  {author} {\bibfnamefont {E.~B.}\ \bibnamefont {Myers}}, \ and\ \bibinfo
  {author} {\bibfnamefont {D.~C.}\ \bibnamefont {Ralph}},\ }\href {\doibase
  10.1103/PhysRevLett.84.3149} {\bibfield  {journal} {\bibinfo  {journal}
  {Physical Review Letters}\ }\textbf {\bibinfo {volume} {84}},\ \bibinfo
  {pages} {3149} (\bibinfo {year} {2000})}\BibitemShut {NoStop}%
\bibitem [{\citenamefont {Mangin}\ \emph {et~al.}(2006)\citenamefont {Mangin},
  \citenamefont {Ravelosona}, \citenamefont {Katine}, \citenamefont {Carey},
  \citenamefont {Terris},\ and\ \citenamefont
  {Fullerton}}]{mangin_current-induced_2006}%
  \BibitemOpen
  \bibfield  {author} {\bibinfo {author} {\bibfnamefont {S.}~\bibnamefont
  {Mangin}}, \bibinfo {author} {\bibfnamefont {D.}~\bibnamefont {Ravelosona}},
  \bibinfo {author} {\bibfnamefont {J.~A.}\ \bibnamefont {Katine}}, \bibinfo
  {author} {\bibfnamefont {M.~J.}\ \bibnamefont {Carey}}, \bibinfo {author}
  {\bibfnamefont {B.~D.}\ \bibnamefont {Terris}}, \ and\ \bibinfo {author}
  {\bibfnamefont {E.~E.}\ \bibnamefont {Fullerton}},\ }\href {\doibase
  10.1038/nmat1595} {\bibfield  {journal} {\bibinfo  {journal} {Nature
  Materials}\ }\textbf {\bibinfo {volume} {5}},\ \bibinfo {pages} {210}
  (\bibinfo {year} {2006})}\BibitemShut {NoStop}%
\bibitem [{\citenamefont {Lide}(2003)}]{lide_CRC_2003}%
  \BibitemOpen
  \bibfield  {author} {\bibinfo {author} {\bibfnamefont {D.~R.}\ \bibnamefont
  {Lide}},\ }\href@noop {} {\emph {\bibinfo {title} {{CRC} {Handbook} of
  {Chemistry} and {Physics}, 84th {Edition}}}}\ (\bibinfo  {publisher} {CRC
  Press},\ \bibinfo {year} {2003})\ \bibinfo {note} {google-Books-ID:
  kTnxSi2B2FcC}\BibitemShut {NoStop}%
\bibitem [{\citenamefont {Nguyen}\ \emph {et~al.}(2016)\citenamefont {Nguyen},
  \citenamefont {Ralph},\ and\ \citenamefont {Buhrman}}]{nguyen_spin_2016}%
  \BibitemOpen
  \bibfield  {author} {\bibinfo {author} {\bibfnamefont {M.-H.}\ \bibnamefont
  {Nguyen}}, \bibinfo {author} {\bibfnamefont {D.}~\bibnamefont {Ralph}}, \
  and\ \bibinfo {author} {\bibfnamefont {R.}~\bibnamefont {Buhrman}},\ }\href
  {\doibase 10.1103/PhysRevLett.116.126601} {\bibfield  {journal} {\bibinfo
  {journal} {Physical Review Letters}\ }\textbf {\bibinfo {volume} {116}},\
  \bibinfo {pages} {126601} (\bibinfo {year} {2016})}\BibitemShut {NoStop}%
\bibitem [{\citenamefont {Wei}\ \emph {et~al.}(2014)\citenamefont {Wei},
  \citenamefont {Obstbaum}, \citenamefont {Ribow}, \citenamefont {Back},\ and\
  \citenamefont {Woltersdorf}}]{wei_spin_2014}%
  \BibitemOpen
  \bibfield  {author} {\bibinfo {author} {\bibfnamefont {D.}~\bibnamefont
  {Wei}}, \bibinfo {author} {\bibfnamefont {M.}~\bibnamefont {Obstbaum}},
  \bibinfo {author} {\bibfnamefont {M.}~\bibnamefont {Ribow}}, \bibinfo
  {author} {\bibfnamefont {C.~H.}\ \bibnamefont {Back}}, \ and\ \bibinfo
  {author} {\bibfnamefont {G.}~\bibnamefont {Woltersdorf}},\ }\href {\doibase
  10.1038/ncomms4768} {\bibfield  {journal} {\bibinfo  {journal} {Nature
  Communications}\ }\textbf {\bibinfo {volume} {5}} (\bibinfo {year} {2014}),\
  10.1038/ncomms4768}\BibitemShut {NoStop}%
\bibitem [{\citenamefont {Hahn}\ \emph {et~al.}(2013)\citenamefont {Hahn},
  \citenamefont {de~Loubens}, \citenamefont {Viret}, \citenamefont {Klein},
  \citenamefont {Naletov},\ and\ \citenamefont
  {Ben~Youssef}}]{hahn_detection_2013}%
  \BibitemOpen
  \bibfield  {author} {\bibinfo {author} {\bibfnamefont {C.}~\bibnamefont
  {Hahn}}, \bibinfo {author} {\bibfnamefont {G.}~\bibnamefont {de~Loubens}},
  \bibinfo {author} {\bibfnamefont {M.}~\bibnamefont {Viret}}, \bibinfo
  {author} {\bibfnamefont {O.}~\bibnamefont {Klein}}, \bibinfo {author}
  {\bibfnamefont {V.~V.}\ \bibnamefont {Naletov}}, \ and\ \bibinfo {author}
  {\bibfnamefont {J.}~\bibnamefont {Ben~Youssef}},\ }\href {\doibase
  10.1103/PhysRevLett.111.217204} {\bibfield  {journal} {\bibinfo  {journal}
  {Physical Review Letters}\ }\textbf {\bibinfo {volume} {111}},\ \bibinfo
  {pages} {217204} (\bibinfo {year} {2013})}\BibitemShut {NoStop}%
\bibitem [{\citenamefont {Silva}\ \emph {et~al.}(2016)\citenamefont {Silva},
  \citenamefont {Nembach}, \citenamefont {Shaw}, \citenamefont {Doyle},
  \citenamefont {Oguz}, \citenamefont {O'Brien},\ and\ \citenamefont
  {Doczy}}]{silva_characterization_2016}%
  \BibitemOpen
  \bibfield  {author} {\bibinfo {author} {\bibfnamefont {T.~J.}\ \bibnamefont
  {Silva}}, \bibinfo {author} {\bibfnamefont {H.~T.}\ \bibnamefont {Nembach}},
  \bibinfo {author} {\bibfnamefont {J.~M.}\ \bibnamefont {Shaw}}, \bibinfo
  {author} {\bibfnamefont {B.}~\bibnamefont {Doyle}}, \bibinfo {author}
  {\bibfnamefont {K.}~\bibnamefont {Oguz}}, \bibinfo {author} {\bibfnamefont
  {K.}~\bibnamefont {O'Brien}}, \ and\ \bibinfo {author} {\bibfnamefont
  {M.}~\bibnamefont {Doczy}},\ }in\ \href {www.panstanford.com} {\emph
  {\bibinfo {booktitle} {Metrology and {Diagnostic} {Techniques} for
  {Nanoelectronics}}}},\ \bibinfo {editor} {edited by\ \bibinfo {editor}
  {\bibfnamefont {Z.}~\bibnamefont {Ma}}\ and\ \bibinfo {editor} {\bibfnamefont
  {D.~G.}\ \bibnamefont {Seiler}}}\ (\bibinfo  {publisher} {Pan Stanford
  Publishing Pte. Ltd},\ \bibinfo {year} {2016})\BibitemShut {NoStop}%
\bibitem [{\citenamefont {Mallinson}(2012)}]{mallinson_foundations_2012}%
  \BibitemOpen
  \bibfield  {author} {\bibinfo {author} {\bibfnamefont {J.~C.}\ \bibnamefont
  {Mallinson}},\ }\href@noop {} {\emph {\bibinfo {title} {The {Foundations} of
  {Magnetic} {Recording}}}}\ (\bibinfo  {publisher} {Academic Press},\ \bibinfo
  {year} {2012})\ \bibinfo {note} {google-Books-ID: wIXIy0hIAHoC}\BibitemShut
  {NoStop}%
\bibitem [{\citenamefont {Rosa}(1908)}]{rosa_self_1908}%
  \BibitemOpen
  \bibfield  {author} {\bibinfo {author} {\bibfnamefont {E.~B.}\ \bibnamefont
  {Rosa}},\ }\href@noop {} {\emph {\bibinfo {title} {The self and mutual
  inductances of linear conductors}}}\ (\bibinfo  {publisher} {US Department of
  Commerce and Labor, Bureau of Standards},\ \bibinfo {year}
  {1908})\BibitemShut {NoStop}%
\bibitem [{\citenamefont {Schneider}\ \emph {et~al.}(2007)\citenamefont
  {Schneider}, \citenamefont {Shaw}, \citenamefont {Kos}, \citenamefont
  {Gerrits}, \citenamefont {Silva},\ and\ \citenamefont
  {McMichael}}]{schneider_spin_2007}%
  \BibitemOpen
  \bibfield  {author} {\bibinfo {author} {\bibfnamefont {M.~L.}\ \bibnamefont
  {Schneider}}, \bibinfo {author} {\bibfnamefont {J.~M.}\ \bibnamefont {Shaw}},
  \bibinfo {author} {\bibfnamefont {A.~B.}\ \bibnamefont {Kos}}, \bibinfo
  {author} {\bibfnamefont {T.}~\bibnamefont {Gerrits}}, \bibinfo {author}
  {\bibfnamefont {T.~J.}\ \bibnamefont {Silva}}, \ and\ \bibinfo {author}
  {\bibfnamefont {R.~D.}\ \bibnamefont {McMichael}},\ }\href {\doibase
  10.1063/1.2812541} {\bibfield  {journal} {\bibinfo  {journal} {Journal of
  Applied Physics}\ }\textbf {\bibinfo {volume} {102}},\ \bibinfo {pages}
  {103909} (\bibinfo {year} {2007})}\BibitemShut {NoStop}%
\bibitem [{Note1()}]{Note1}%
  \BibitemOpen
  \bibinfo {note} {Python code to execute this analysis is shared publicly at
  \url{https://github.com/berger156/VNA\_FMR}.}\BibitemShut {Stop}%
\bibitem [{\citenamefont {Niimi}\ \emph {et~al.}(2011)\citenamefont {Niimi},
  \citenamefont {Morota}, \citenamefont {Wei}, \citenamefont {Deranlot},
  \citenamefont {Basletic}, \citenamefont {Hamzic}, \citenamefont {Fert},\ and\
  \citenamefont {Otani}}]{niimi_extrinsic_2011}%
  \BibitemOpen
  \bibfield  {author} {\bibinfo {author} {\bibfnamefont {Y.}~\bibnamefont
  {Niimi}}, \bibinfo {author} {\bibfnamefont {M.}~\bibnamefont {Morota}},
  \bibinfo {author} {\bibfnamefont {D.~H.}\ \bibnamefont {Wei}}, \bibinfo
  {author} {\bibfnamefont {C.}~\bibnamefont {Deranlot}}, \bibinfo {author}
  {\bibfnamefont {M.}~\bibnamefont {Basletic}}, \bibinfo {author}
  {\bibfnamefont {A.}~\bibnamefont {Hamzic}}, \bibinfo {author} {\bibfnamefont
  {A.}~\bibnamefont {Fert}}, \ and\ \bibinfo {author} {\bibfnamefont
  {Y.}~\bibnamefont {Otani}},\ }\href {\doibase 10.1103/PhysRevLett.106.126601}
  {\bibfield  {journal} {\bibinfo  {journal} {Physical Review Letters}\
  }\textbf {\bibinfo {volume} {106}},\ \bibinfo {pages} {126601} (\bibinfo
  {year} {2011})}\BibitemShut {NoStop}%
\bibitem [{\citenamefont {Sinova}\ \emph {et~al.}(2015)\citenamefont {Sinova},
  \citenamefont {Valenzuela}, \citenamefont {Wunderlich}, \citenamefont
  {Back},\ and\ \citenamefont {Jungwirth}}]{sinova_spin_2015}%
  \BibitemOpen
  \bibfield  {author} {\bibinfo {author} {\bibfnamefont {J.}~\bibnamefont
  {Sinova}}, \bibinfo {author} {\bibfnamefont {S.~O.}\ \bibnamefont
  {Valenzuela}}, \bibinfo {author} {\bibfnamefont {J.}~\bibnamefont
  {Wunderlich}}, \bibinfo {author} {\bibfnamefont {C.}~\bibnamefont {Back}}, \
  and\ \bibinfo {author} {\bibfnamefont {T.}~\bibnamefont {Jungwirth}},\ }\href
  {\doibase 10.1103/RevModPhys.87.1213} {\bibfield  {journal} {\bibinfo
  {journal} {Reviews of Modern Physics}\ }\textbf {\bibinfo {volume} {87}},\
  \bibinfo {pages} {1213} (\bibinfo {year} {2015})}\BibitemShut {NoStop}%
\bibitem [{\citenamefont {Wessel-Berg}\ and\ \citenamefont
  {Bertram}(1978)}]{wessel-berg_generalized_1978}%
  \BibitemOpen
  \bibfield  {author} {\bibinfo {author} {\bibfnamefont {T.}~\bibnamefont
  {Wessel-Berg}}\ and\ \bibinfo {author} {\bibfnamefont {H.}~\bibnamefont
  {Bertram}},\ }\href {\doibase 10.1109/TMAG.1978.1059737} {\bibfield
  {journal} {\bibinfo  {journal} {IEEE Transactions on Magnetics}\ }\textbf
  {\bibinfo {volume} {14}},\ \bibinfo {pages} {129} (\bibinfo {year}
  {1978})}\BibitemShut {NoStop}%
\bibitem [{\citenamefont {Kim}\ \emph {et~al.}(2013)\citenamefont {Kim},
  \citenamefont {Lee}, \citenamefont {Lee},\ and\ \citenamefont
  {Stiles}}]{kim_chirality_2013}%
  \BibitemOpen
  \bibfield  {author} {\bibinfo {author} {\bibfnamefont {K.-W.}\ \bibnamefont
  {Kim}}, \bibinfo {author} {\bibfnamefont {H.-W.}\ \bibnamefont {Lee}},
  \bibinfo {author} {\bibfnamefont {K.-J.}\ \bibnamefont {Lee}}, \ and\
  \bibinfo {author} {\bibfnamefont {M.~D.}\ \bibnamefont {Stiles}},\ }\href
  {\doibase 10.1103/PhysRevLett.111.216601} {\bibfield  {journal} {\bibinfo
  {journal} {Physical Review Letters}\ }\textbf {\bibinfo {volume} {111}},\
  \bibinfo {pages} {216601} (\bibinfo {year} {2013})}\BibitemShut {NoStop}%
\bibitem [{\citenamefont {Zhu}\ \emph {et~al.}(2010)\citenamefont {Zhu},
  \citenamefont {Dennis},\ and\ \citenamefont
  {McMichael}}]{zhu_temperature_2010}%
  \BibitemOpen
  \bibfield  {author} {\bibinfo {author} {\bibfnamefont {M.}~\bibnamefont
  {Zhu}}, \bibinfo {author} {\bibfnamefont {C.~L.}\ \bibnamefont {Dennis}}, \
  and\ \bibinfo {author} {\bibfnamefont {R.~D.}\ \bibnamefont {McMichael}},\
  }\href {\doibase 10.1103/PhysRevB.81.140407} {\bibfield  {journal} {\bibinfo
  {journal} {Physical Review B}\ }\textbf {\bibinfo {volume} {81}},\ \bibinfo
  {pages} {140407} (\bibinfo {year} {2010})}\BibitemShut {NoStop}%
\bibitem [{\citenamefont {Haney}\ \emph
  {et~al.}(2013{\natexlab{a}})\citenamefont {Haney}, \citenamefont {Lee},
  \citenamefont {Lee}, \citenamefont {Manchon},\ and\ \citenamefont
  {Stiles}}]{haney_current-induced_2013}%
  \BibitemOpen
  \bibfield  {author} {\bibinfo {author} {\bibfnamefont {P.~M.}\ \bibnamefont
  {Haney}}, \bibinfo {author} {\bibfnamefont {H.-W.}\ \bibnamefont {Lee}},
  \bibinfo {author} {\bibfnamefont {K.-J.}\ \bibnamefont {Lee}}, \bibinfo
  {author} {\bibfnamefont {A.}~\bibnamefont {Manchon}}, \ and\ \bibinfo
  {author} {\bibfnamefont {M.~D.}\ \bibnamefont {Stiles}},\ }\href {\doibase
  10.1103/PhysRevB.88.214417} {\bibfield  {journal} {\bibinfo  {journal}
  {Physical Review B}\ }\textbf {\bibinfo {volume} {88}},\ \bibinfo {pages}
  {214417} (\bibinfo {year} {2013}{\natexlab{a}})}\BibitemShut {NoStop}%
\bibitem [{\citenamefont {Cercellier}\ \emph {et~al.}(2006)\citenamefont
  {Cercellier}, \citenamefont {Didiot}, \citenamefont {Fagot-Revurat},
  \citenamefont {Kierren}, \citenamefont {Moreau}, \citenamefont {Malterre},\
  and\ \citenamefont {Reinert}}]{cercellier_interplay_2006}%
  \BibitemOpen
  \bibfield  {author} {\bibinfo {author} {\bibfnamefont {H.}~\bibnamefont
  {Cercellier}}, \bibinfo {author} {\bibfnamefont {C.}~\bibnamefont {Didiot}},
  \bibinfo {author} {\bibfnamefont {Y.}~\bibnamefont {Fagot-Revurat}}, \bibinfo
  {author} {\bibfnamefont {B.}~\bibnamefont {Kierren}}, \bibinfo {author}
  {\bibfnamefont {L.}~\bibnamefont {Moreau}}, \bibinfo {author} {\bibfnamefont
  {D.}~\bibnamefont {Malterre}}, \ and\ \bibinfo {author} {\bibfnamefont
  {F.}~\bibnamefont {Reinert}},\ }\href {\doibase 10.1103/PhysRevB.73.195413}
  {\bibfield  {journal} {\bibinfo  {journal} {Physical Review B}\ }\textbf
  {\bibinfo {volume} {73}},\ \bibinfo {pages} {195413} (\bibinfo {year}
  {2006})}\BibitemShut {NoStop}%
\bibitem [{\citenamefont {Koroteev}\ \emph {et~al.}(2004)\citenamefont
  {Koroteev}, \citenamefont {Bihlmayer}, \citenamefont {Gayone}, \citenamefont
  {Chulkov}, \citenamefont {Blügel}, \citenamefont {Echenique},\ and\
  \citenamefont {Hofmann}}]{koroteev_strong_2004}%
  \BibitemOpen
  \bibfield  {author} {\bibinfo {author} {\bibfnamefont {Y.~M.}\ \bibnamefont
  {Koroteev}}, \bibinfo {author} {\bibfnamefont {G.}~\bibnamefont {Bihlmayer}},
  \bibinfo {author} {\bibfnamefont {J.~E.}\ \bibnamefont {Gayone}}, \bibinfo
  {author} {\bibfnamefont {E.~V.}\ \bibnamefont {Chulkov}}, \bibinfo {author}
  {\bibfnamefont {S.}~\bibnamefont {Blügel}}, \bibinfo {author} {\bibfnamefont
  {P.~M.}\ \bibnamefont {Echenique}}, \ and\ \bibinfo {author} {\bibfnamefont
  {P.}~\bibnamefont {Hofmann}},\ }\href {\doibase
  10.1103/PhysRevLett.93.046403} {\bibfield  {journal} {\bibinfo  {journal}
  {Physical Review Letters}\ }\textbf {\bibinfo {volume} {93}},\ \bibinfo
  {pages} {046403} (\bibinfo {year} {2004})}\BibitemShut {NoStop}%
\bibitem [{\citenamefont {Yaji}\ \emph {et~al.}(2010)\citenamefont {Yaji},
  \citenamefont {Ohtsubo}, \citenamefont {Hatta}, \citenamefont {Okuyama},
  \citenamefont {Miyamoto}, \citenamefont {Okuda}, \citenamefont {Kimura},
  \citenamefont {Namatame}, \citenamefont {Taniguchi},\ and\ \citenamefont
  {Aruga}}]{yaji_large_2010}%
  \BibitemOpen
  \bibfield  {author} {\bibinfo {author} {\bibfnamefont {K.}~\bibnamefont
  {Yaji}}, \bibinfo {author} {\bibfnamefont {Y.}~\bibnamefont {Ohtsubo}},
  \bibinfo {author} {\bibfnamefont {S.}~\bibnamefont {Hatta}}, \bibinfo
  {author} {\bibfnamefont {H.}~\bibnamefont {Okuyama}}, \bibinfo {author}
  {\bibfnamefont {K.}~\bibnamefont {Miyamoto}}, \bibinfo {author}
  {\bibfnamefont {T.}~\bibnamefont {Okuda}}, \bibinfo {author} {\bibfnamefont
  {A.}~\bibnamefont {Kimura}}, \bibinfo {author} {\bibfnamefont
  {H.}~\bibnamefont {Namatame}}, \bibinfo {author} {\bibfnamefont
  {M.}~\bibnamefont {Taniguchi}}, \ and\ \bibinfo {author} {\bibfnamefont
  {T.}~\bibnamefont {Aruga}},\ }\href {\doibase 10.1038/ncomms1016} {\bibfield
  {journal} {\bibinfo  {journal} {Nature Communications}\ }\textbf {\bibinfo
  {volume} {1}},\ \bibinfo {pages} {ncomms1016} (\bibinfo {year}
  {2010})}\BibitemShut {NoStop}%
\bibitem [{\citenamefont {Ast}\ \emph {et~al.}(2007)\citenamefont {Ast},
  \citenamefont {Henk}, \citenamefont {Ernst}, \citenamefont {Moreschini},
  \citenamefont {Falub}, \citenamefont {Pacilé}, \citenamefont {Bruno},
  \citenamefont {Kern},\ and\ \citenamefont {Grioni}}]{ast_giant_2007}%
  \BibitemOpen
  \bibfield  {author} {\bibinfo {author} {\bibfnamefont {C.~R.}\ \bibnamefont
  {Ast}}, \bibinfo {author} {\bibfnamefont {J.}~\bibnamefont {Henk}}, \bibinfo
  {author} {\bibfnamefont {A.}~\bibnamefont {Ernst}}, \bibinfo {author}
  {\bibfnamefont {L.}~\bibnamefont {Moreschini}}, \bibinfo {author}
  {\bibfnamefont {M.~C.}\ \bibnamefont {Falub}}, \bibinfo {author}
  {\bibfnamefont {D.}~\bibnamefont {Pacilé}}, \bibinfo {author} {\bibfnamefont
  {P.}~\bibnamefont {Bruno}}, \bibinfo {author} {\bibfnamefont
  {K.}~\bibnamefont {Kern}}, \ and\ \bibinfo {author} {\bibfnamefont
  {M.}~\bibnamefont {Grioni}},\ }\href {\doibase 10.1103/PhysRevLett.98.186807}
  {\bibfield  {journal} {\bibinfo  {journal} {Physical Review Letters}\
  }\textbf {\bibinfo {volume} {98}},\ \bibinfo {pages} {186807} (\bibinfo
  {year} {2007})}\BibitemShut {NoStop}%
\bibitem [{\citenamefont {Nembach}\ \emph {et~al.}(2015)\citenamefont
  {Nembach}, \citenamefont {Shaw}, \citenamefont {Weiler}, \citenamefont
  {Jué},\ and\ \citenamefont {Silva}}]{nembach_linear_2015}%
  \BibitemOpen
  \bibfield  {author} {\bibinfo {author} {\bibfnamefont {H.~T.}\ \bibnamefont
  {Nembach}}, \bibinfo {author} {\bibfnamefont {J.~M.}\ \bibnamefont {Shaw}},
  \bibinfo {author} {\bibfnamefont {M.}~\bibnamefont {Weiler}}, \bibinfo
  {author} {\bibfnamefont {E.}~\bibnamefont {Jué}}, \ and\ \bibinfo {author}
  {\bibfnamefont {T.~J.}\ \bibnamefont {Silva}},\ }\href {\doibase
  10.1038/nphys3418} {\bibfield  {journal} {\bibinfo  {journal} {Nature
  Physics}\ }\textbf {\bibinfo {volume} {11}},\ \bibinfo {pages} {825}
  (\bibinfo {year} {2015})}\BibitemShut {NoStop}%
\bibitem [{\citenamefont {Boone}\ \emph {et~al.}(2013)\citenamefont {Boone},
  \citenamefont {Nembach}, \citenamefont {Shaw},\ and\ \citenamefont
  {Silva}}]{boone_spin_2013}%
  \BibitemOpen
  \bibfield  {author} {\bibinfo {author} {\bibfnamefont {C.~T.}\ \bibnamefont
  {Boone}}, \bibinfo {author} {\bibfnamefont {H.~T.}\ \bibnamefont {Nembach}},
  \bibinfo {author} {\bibfnamefont {J.~M.}\ \bibnamefont {Shaw}}, \ and\
  \bibinfo {author} {\bibfnamefont {T.~J.}\ \bibnamefont {Silva}},\ }\href
  {\doibase 10.1063/1.4801799} {\bibfield  {journal} {\bibinfo  {journal}
  {Journal of Applied Physics}\ }\textbf {\bibinfo {volume} {113}},\ \bibinfo
  {pages} {153906} (\bibinfo {year} {2013})}\BibitemShut {NoStop}%
\bibitem [{\citenamefont {Rojas-Sánchez}\ \emph {et~al.}(2014)\citenamefont
  {Rojas-Sánchez}, \citenamefont {Reyren}, \citenamefont {Laczkowski},
  \citenamefont {Savero}, \citenamefont {Attané}, \citenamefont {Deranlot},
  \citenamefont {Jamet}, \citenamefont {George}, \citenamefont {Vila},\ and\
  \citenamefont {Jaffrès}}]{rojas-sanchez_spin_2014}%
  \BibitemOpen
  \bibfield  {author} {\bibinfo {author} {\bibfnamefont {J.-C.}\ \bibnamefont
  {Rojas-Sánchez}}, \bibinfo {author} {\bibfnamefont {N.}~\bibnamefont
  {Reyren}}, \bibinfo {author} {\bibfnamefont {P.}~\bibnamefont {Laczkowski}},
  \bibinfo {author} {\bibfnamefont {W.}~\bibnamefont {Savero}}, \bibinfo
  {author} {\bibfnamefont {J.-P.}\ \bibnamefont {Attané}}, \bibinfo {author}
  {\bibfnamefont {C.}~\bibnamefont {Deranlot}}, \bibinfo {author}
  {\bibfnamefont {M.}~\bibnamefont {Jamet}}, \bibinfo {author} {\bibfnamefont
  {J.-M.}\ \bibnamefont {George}}, \bibinfo {author} {\bibfnamefont
  {L.}~\bibnamefont {Vila}}, \ and\ \bibinfo {author} {\bibfnamefont
  {H.}~\bibnamefont {Jaffrès}},\ }\href {\doibase
  10.1103/PhysRevLett.112.106602} {\bibfield  {journal} {\bibinfo  {journal}
  {Physical Review Letters}\ }\textbf {\bibinfo {volume} {112}},\ \bibinfo
  {pages} {106602} (\bibinfo {year} {2014})}\BibitemShut {NoStop}%
\bibitem [{\citenamefont {Caminale}\ \emph {et~al.}(2016)\citenamefont
  {Caminale}, \citenamefont {Ghosh}, \citenamefont {Auffret}, \citenamefont
  {Ebels}, \citenamefont {Ollefs}, \citenamefont {Wilhelm}, \citenamefont
  {Rogalev},\ and\ \citenamefont {Bailey}}]{caminale_spin_2016}%
  \BibitemOpen
  \bibfield  {author} {\bibinfo {author} {\bibfnamefont {M.}~\bibnamefont
  {Caminale}}, \bibinfo {author} {\bibfnamefont {A.}~\bibnamefont {Ghosh}},
  \bibinfo {author} {\bibfnamefont {S.}~\bibnamefont {Auffret}}, \bibinfo
  {author} {\bibfnamefont {U.}~\bibnamefont {Ebels}}, \bibinfo {author}
  {\bibfnamefont {K.}~\bibnamefont {Ollefs}}, \bibinfo {author} {\bibfnamefont
  {F.}~\bibnamefont {Wilhelm}}, \bibinfo {author} {\bibfnamefont
  {A.}~\bibnamefont {Rogalev}}, \ and\ \bibinfo {author} {\bibfnamefont
  {W.~E.}\ \bibnamefont {Bailey}},\ }\href {\doibase
  10.1103/PhysRevB.94.014414} {\bibfield  {journal} {\bibinfo  {journal}
  {Physical Review B}\ }\textbf {\bibinfo {volume} {94}},\ \bibinfo {pages}
  {014414} (\bibinfo {year} {2016})}\BibitemShut {NoStop}%
\bibitem [{\citenamefont {Haney}\ \emph
  {et~al.}(2013{\natexlab{b}})\citenamefont {Haney}, \citenamefont {Lee},
  \citenamefont {Lee}, \citenamefont {Manchon},\ and\ \citenamefont
  {Stiles}}]{haney_current_2013}%
  \BibitemOpen
  \bibfield  {author} {\bibinfo {author} {\bibfnamefont {P.~M.}\ \bibnamefont
  {Haney}}, \bibinfo {author} {\bibfnamefont {H.-W.}\ \bibnamefont {Lee}},
  \bibinfo {author} {\bibfnamefont {K.-J.}\ \bibnamefont {Lee}}, \bibinfo
  {author} {\bibfnamefont {A.}~\bibnamefont {Manchon}}, \ and\ \bibinfo
  {author} {\bibfnamefont {M.~D.}\ \bibnamefont {Stiles}},\ }\href {\doibase
  10.1103/PhysRevB.87.174411} {\bibfield  {journal} {\bibinfo  {journal}
  {Physical Review B}\ }\textbf {\bibinfo {volume} {87}},\ \bibinfo {pages}
  {174411} (\bibinfo {year} {2013}{\natexlab{b}})}\BibitemShut {NoStop}%
\bibitem [{\citenamefont {Isasa}\ \emph {et~al.}(2015)\citenamefont {Isasa},
  \citenamefont {Villamor}, \citenamefont {Hueso}, \citenamefont {Gradhand},\
  and\ \citenamefont {Casanova}}]{isasa_temperature_2015}%
  \BibitemOpen
  \bibfield  {author} {\bibinfo {author} {\bibfnamefont {M.}~\bibnamefont
  {Isasa}}, \bibinfo {author} {\bibfnamefont {E.}~\bibnamefont {Villamor}},
  \bibinfo {author} {\bibfnamefont {L.~E.}\ \bibnamefont {Hueso}}, \bibinfo
  {author} {\bibfnamefont {M.}~\bibnamefont {Gradhand}}, \ and\ \bibinfo
  {author} {\bibfnamefont {F.}~\bibnamefont {Casanova}},\ }\href {\doibase
  10.1103/PhysRevB.91.024402} {\bibfield  {journal} {\bibinfo  {journal}
  {Physical Review B}\ }\textbf {\bibinfo {volume} {91}},\ \bibinfo {pages}
  {024402} (\bibinfo {year} {2015})}\BibitemShut {NoStop}%
\bibitem [{\citenamefont {Mosendz}\ \emph {et~al.}(2010)\citenamefont
  {Mosendz}, \citenamefont {Vlaminck}, \citenamefont {Pearson}, \citenamefont
  {Fradin}, \citenamefont {Bauer}, \citenamefont {Bader},\ and\ \citenamefont
  {Hoffmann}}]{mosendz_detection_2010}%
  \BibitemOpen
  \bibfield  {author} {\bibinfo {author} {\bibfnamefont {O.}~\bibnamefont
  {Mosendz}}, \bibinfo {author} {\bibfnamefont {V.}~\bibnamefont {Vlaminck}},
  \bibinfo {author} {\bibfnamefont {J.~E.}\ \bibnamefont {Pearson}}, \bibinfo
  {author} {\bibfnamefont {F.~Y.}\ \bibnamefont {Fradin}}, \bibinfo {author}
  {\bibfnamefont {G.~E.~W.}\ \bibnamefont {Bauer}}, \bibinfo {author}
  {\bibfnamefont {S.~D.}\ \bibnamefont {Bader}}, \ and\ \bibinfo {author}
  {\bibfnamefont {A.}~\bibnamefont {Hoffmann}},\ }\href {\doibase
  10.1103/PhysRevB.82.214403} {\bibfield  {journal} {\bibinfo  {journal}
  {Physical Review B}\ }\textbf {\bibinfo {volume} {82}},\ \bibinfo {pages}
  {214403} (\bibinfo {year} {2010})}\BibitemShut {NoStop}%
\bibitem [{\citenamefont {Morota}\ \emph {et~al.}(2011)\citenamefont {Morota},
  \citenamefont {Niimi}, \citenamefont {Ohnishi}, \citenamefont {Wei},
  \citenamefont {Tanaka}, \citenamefont {Kontani}, \citenamefont {Kimura},\
  and\ \citenamefont {Otani}}]{morota_indication_2011}%
  \BibitemOpen
  \bibfield  {author} {\bibinfo {author} {\bibfnamefont {M.}~\bibnamefont
  {Morota}}, \bibinfo {author} {\bibfnamefont {Y.}~\bibnamefont {Niimi}},
  \bibinfo {author} {\bibfnamefont {K.}~\bibnamefont {Ohnishi}}, \bibinfo
  {author} {\bibfnamefont {D.~H.}\ \bibnamefont {Wei}}, \bibinfo {author}
  {\bibfnamefont {T.}~\bibnamefont {Tanaka}}, \bibinfo {author} {\bibfnamefont
  {H.}~\bibnamefont {Kontani}}, \bibinfo {author} {\bibfnamefont
  {T.}~\bibnamefont {Kimura}}, \ and\ \bibinfo {author} {\bibfnamefont
  {Y.}~\bibnamefont {Otani}},\ }\href {\doibase 10.1103/PhysRevB.83.174405}
  {\bibfield  {journal} {\bibinfo  {journal} {Physical Review B}\ }\textbf
  {\bibinfo {volume} {83}},\ \bibinfo {pages} {174405} (\bibinfo {year}
  {2011})}\BibitemShut {NoStop}%
\bibitem [{\citenamefont {Liu}\ \emph {et~al.}(2011)\citenamefont {Liu},
  \citenamefont {Moriyama}, \citenamefont {Ralph},\ and\ \citenamefont
  {Buhrman}}]{liu_spin-torque_2011}%
  \BibitemOpen
  \bibfield  {author} {\bibinfo {author} {\bibfnamefont {L.}~\bibnamefont
  {Liu}}, \bibinfo {author} {\bibfnamefont {T.}~\bibnamefont {Moriyama}},
  \bibinfo {author} {\bibfnamefont {D.~C.}\ \bibnamefont {Ralph}}, \ and\
  \bibinfo {author} {\bibfnamefont {R.~A.}\ \bibnamefont {Buhrman}},\ }\href
  {\doibase 10.1103/PhysRevLett.106.036601} {\bibfield  {journal} {\bibinfo
  {journal} {Physical Review Letters}\ }\textbf {\bibinfo {volume} {106}},\
  \bibinfo {pages} {036601} (\bibinfo {year} {2011})}\BibitemShut {NoStop}%
\bibitem [{\citenamefont {Weiler}\ \emph {et~al.}(2013)\citenamefont {Weiler},
  \citenamefont {Althammer}, \citenamefont {Schreier}, \citenamefont {Lotze},
  \citenamefont {Pernpeintner}, \citenamefont {Meyer}, \citenamefont {Huebl},
  \citenamefont {Gross}, \citenamefont {Kamra}, \citenamefont {Xiao},
  \citenamefont {Chen}, \citenamefont {Jiao}, \citenamefont {Bauer},\ and\
  \citenamefont {Goennenwein}}]{weiler_experimental_2013}%
  \BibitemOpen
  \bibfield  {author} {\bibinfo {author} {\bibfnamefont {M.}~\bibnamefont
  {Weiler}}, \bibinfo {author} {\bibfnamefont {M.}~\bibnamefont {Althammer}},
  \bibinfo {author} {\bibfnamefont {M.}~\bibnamefont {Schreier}}, \bibinfo
  {author} {\bibfnamefont {J.}~\bibnamefont {Lotze}}, \bibinfo {author}
  {\bibfnamefont {M.}~\bibnamefont {Pernpeintner}}, \bibinfo {author}
  {\bibfnamefont {S.}~\bibnamefont {Meyer}}, \bibinfo {author} {\bibfnamefont
  {H.}~\bibnamefont {Huebl}}, \bibinfo {author} {\bibfnamefont
  {R.}~\bibnamefont {Gross}}, \bibinfo {author} {\bibfnamefont
  {A.}~\bibnamefont {Kamra}}, \bibinfo {author} {\bibfnamefont
  {J.}~\bibnamefont {Xiao}}, \bibinfo {author} {\bibfnamefont {Y.-T.}\
  \bibnamefont {Chen}}, \bibinfo {author} {\bibfnamefont {H.}~\bibnamefont
  {Jiao}}, \bibinfo {author} {\bibfnamefont {G.~E.~W.}\ \bibnamefont {Bauer}},
  \ and\ \bibinfo {author} {\bibfnamefont {S.~T.~B.}\ \bibnamefont
  {Goennenwein}},\ }\href {\doibase 10.1103/PhysRevLett.111.176601} {\bibfield
  {journal} {\bibinfo  {journal} {Physical Review Letters}\ }\textbf {\bibinfo
  {volume} {111}},\ \bibinfo {pages} {176601} (\bibinfo {year}
  {2013})}\BibitemShut {NoStop}%
\bibitem [{\citenamefont {Obstbaum}\ \emph {et~al.}(2014)\citenamefont
  {Obstbaum}, \citenamefont {Härtinger}, \citenamefont {Bauer}, \citenamefont
  {Meier}, \citenamefont {Swientek}, \citenamefont {Back},\ and\ \citenamefont
  {Woltersdorf}}]{obstbaum_inverse_2014}%
  \BibitemOpen
  \bibfield  {author} {\bibinfo {author} {\bibfnamefont {M.}~\bibnamefont
  {Obstbaum}}, \bibinfo {author} {\bibfnamefont {M.}~\bibnamefont
  {Härtinger}}, \bibinfo {author} {\bibfnamefont {H.~G.}\ \bibnamefont
  {Bauer}}, \bibinfo {author} {\bibfnamefont {T.}~\bibnamefont {Meier}},
  \bibinfo {author} {\bibfnamefont {F.}~\bibnamefont {Swientek}}, \bibinfo
  {author} {\bibfnamefont {C.~H.}\ \bibnamefont {Back}}, \ and\ \bibinfo
  {author} {\bibfnamefont {G.}~\bibnamefont {Woltersdorf}},\ }\href {\doibase
  10.1103/PhysRevB.89.060407} {\bibfield  {journal} {\bibinfo  {journal}
  {Physical Review B}\ }\textbf {\bibinfo {volume} {89}},\ \bibinfo {pages}
  {060407} (\bibinfo {year} {2014})}\BibitemShut {NoStop}%
\bibitem [{\citenamefont {Pai}\ \emph {et~al.}(2015)\citenamefont {Pai},
  \citenamefont {Ou}, \citenamefont {Vilela-Leão}, \citenamefont {Ralph},\
  and\ \citenamefont {Buhrman}}]{pai_dependence_2015}%
  \BibitemOpen
  \bibfield  {author} {\bibinfo {author} {\bibfnamefont {C.-F.}\ \bibnamefont
  {Pai}}, \bibinfo {author} {\bibfnamefont {Y.}~\bibnamefont {Ou}}, \bibinfo
  {author} {\bibfnamefont {L.~H.}\ \bibnamefont {Vilela-Leão}}, \bibinfo
  {author} {\bibfnamefont {D.~C.}\ \bibnamefont {Ralph}}, \ and\ \bibinfo
  {author} {\bibfnamefont {R.~A.}\ \bibnamefont {Buhrman}},\ }\href {\doibase
  10.1103/PhysRevB.92.064426} {\bibfield  {journal} {\bibinfo  {journal}
  {Physical Review B}\ }\textbf {\bibinfo {volume} {92}},\ \bibinfo {pages}
  {064426} (\bibinfo {year} {2015})}\BibitemShut {NoStop}%
\bibitem [{\citenamefont {Liu}\ \emph {et~al.}(2014)\citenamefont {Liu},
  \citenamefont {Yuan}, \citenamefont {Wesselink}, \citenamefont {Starikov},\
  and\ \citenamefont {Kelly}}]{liu_interface_2014}%
  \BibitemOpen
  \bibfield  {author} {\bibinfo {author} {\bibfnamefont {Y.}~\bibnamefont
  {Liu}}, \bibinfo {author} {\bibfnamefont {Z.}~\bibnamefont {Yuan}}, \bibinfo
  {author} {\bibfnamefont {R.}~\bibnamefont {Wesselink}}, \bibinfo {author}
  {\bibfnamefont {A.~A.}\ \bibnamefont {Starikov}}, \ and\ \bibinfo {author}
  {\bibfnamefont {P.~J.}\ \bibnamefont {Kelly}},\ }\href {\doibase
  10.1103/PhysRevLett.113.207202} {\bibfield  {journal} {\bibinfo  {journal}
  {Physical Review Letters}\ }\textbf {\bibinfo {volume} {113}},\ \bibinfo
  {pages} {207202} (\bibinfo {year} {2014})}\BibitemShut {NoStop}%
\end{thebibliography}%


\begin{thebibliography}{2}%
\makeatletter
\providecommand \@ifxundefined [1]{%
 \@ifx{#1\undefined}
}%
\providecommand \@ifnum [1]{%
 \ifnum #1\expandafter \@firstoftwo
 \else \expandafter \@secondoftwo
 \fi
}%
\providecommand \@ifx [1]{%
 \ifx #1\expandafter \@firstoftwo
 \else \expandafter \@secondoftwo
 \fi
}%
\providecommand \natexlab [1]{#1}%
\providecommand \enquote  [1]{``#1''}%
\providecommand \bibnamefont  [1]{#1}%
\providecommand \bibfnamefont [1]{#1}%
\providecommand \citenamefont [1]{#1}%
\providecommand \href@noop [0]{\@secondoftwo}%
\providecommand \href [0]{\begingroup \@sanitize@url \@href}%
\providecommand \@href[1]{\@@startlink{#1}\@@href}%
\providecommand \@@href[1]{\endgroup#1\@@endlink}%
\providecommand \@sanitize@url [0]{\catcode `\\12\catcode `\$12\catcode
  `\&12\catcode `\#12\catcode `\^12\catcode `\_12\catcode `\%12\relax}%
\providecommand \@@startlink[1]{}%
\providecommand \@@endlink[0]{}%
\providecommand \url  [0]{\begingroup\@sanitize@url \@url }%
\providecommand \@url [1]{\endgroup\@href {#1}{\urlprefix }}%
\providecommand \urlprefix  [0]{URL }%
\providecommand \Eprint [0]{\href }%
\providecommand \doibase [0]{http://dx.doi.org/}%
\providecommand \selectlanguage [0]{\@gobble}%
\providecommand \bibinfo  [0]{\@secondoftwo}%
\providecommand \bibfield  [0]{\@secondoftwo}%
\providecommand \translation [1]{[#1]}%
\providecommand \BibitemOpen [0]{}%
\providecommand \bibitemStop [0]{}%
\providecommand \bibitemNoStop [0]{.\EOS\space}%
\providecommand \EOS [0]{\spacefactor3000\relax}%
\providecommand \BibitemShut  [1]{\csname bibitem#1\endcsname}%
\let\auto@bib@innerbib\@empty
\bibitem [{\citenamefont {Kittel}(2004)}]{kittel_introduction_2004}%
  \BibitemOpen
  \bibfield  {author} {\bibinfo {author} {\bibfnamefont {C.}~\bibnamefont
  {Kittel}},\ }\href@noop {} {\emph {\bibinfo {title} {Introduction to {Solid}
  {State} {Physics}}}},\ \bibinfo {edition} {8th}\ ed.\ (\bibinfo  {publisher}
  {Wiley},\ \bibinfo {year} {2004})\BibitemShut {NoStop}%
\bibitem [{\citenamefont {Jiao}\ and\ \citenamefont
  {Bauer}(2013)}]{jiao_spin_2013}%
  \BibitemOpen
  \bibfield  {author} {\bibinfo {author} {\bibfnamefont {H.}~\bibnamefont
  {Jiao}}\ and\ \bibinfo {author} {\bibfnamefont {G.~E.~W.}\ \bibnamefont
  {Bauer}},\ }\href {\doibase 10.1103/PhysRevLett.110.217602} {\bibfield
  {journal} {\bibinfo  {journal} {Physical Review Letters}\ }\textbf {\bibinfo
  {volume} {110}},\ \bibinfo {pages} {217602} (\bibinfo {year}
  {2013})}\BibitemShut {NoStop}%
\end{thebibliography}%
